\documentclass[12pt,preprint]{aastex}
\usepackage{multirow}
\usepackage{booktabs}
\usepackage{graphicx}
\usepackage{amsmath}
\usepackage{longtable}
\usepackage{rotating}
\usepackage[caption=false]{subfig}
\usepackage{lscape}
\usepackage{color}

\newcommand{\msun}{M_\odot}
\newcommand{\mj}{M_{\rm J}}

\newcommand{\mdisk}{M_{\rm disk}}
\newcommand{\mstar}{M_\star}
\newcommand{\sigmag}{\Sigma_{\rm gas}}
\newcommand{\rhog}{\rho_{\rm gas}}
\newcommand{\rhod}{\rho_{\rm grains}}
\newcommand{\rhosg}{\rho_{\rm small\ grains}}

\newcommand{\rsub}{r_{\rm sub}}
\newcommand{\rin}{r_{\rm in}}
\newcommand{\rout}{r_{\rm out}}
\newcommand{\tirr}{T_{\rm irr}}
\newcommand{\thydro}{T_{\rm hydro}}
\newcommand{\tmcrt}{T_{\rm MCRT}}
\newcommand{\tmcrtmid}{T_{\rm MCRT,midplane}}

\newcommand{\bl}[1]{\mbox{\boldmath$ #1 $}}

\begin{document}
\title{Signatures of Gravitational Instability in Resolved Images of Protostellar Disks}

\shortauthors{Dong et al.}

\author{Ruobing Dong\altaffilmark{1,2,7}, Eduard Vorobyov\altaffilmark{3,4}, Yaroslav Pavlyuchenkov\altaffilmark{5}, Eugene Chiang\altaffilmark{2} \& Hauyu Baobab Liu\altaffilmark{6,8}}

\altaffiltext{1}{Lawrence Berkeley National Lab, Berkeley, CA 94720; rdong2013@berkeley.edu}
\altaffiltext{2}{Department of Astronomy, University of California at Berkeley, Berkeley, CA 94720}
\altaffiltext{3}{Department of Astrophysics, The University of Vienna, Vienna, 1180, Austria}
\altaffiltext{4}{Research Institute of Physics, Southern Federal University, Stachki 194, Rostov-on-Don, 344090, Russia}
\altaffiltext{5}{Institute of Astronomy,  Russian Academy of Sciences, Moscow, Russia}
\altaffiltext{6}{European Southern Observatory (ESO), Karl-Schwarzschild
-Str. 2, D-85748 Garching, Germany}
\altaffiltext{7}{NASA Hubble Fellow}
\altaffiltext{8}{ESO Fellow}

\clearpage

\begin{abstract}

Protostellar (class 0/I) disks, having masses comparable to those
of their nascent host stars, and fed continuously from their natal
infalling envelopes, are prone to gravitational instability (GI).
Motivated by advances in near-infrared (NIR) adaptive
optics imaging and mm-wave interferometry, we explore the 
observational signatures of GI in disks,
using hydrodynamical and Monte Carlo radiative transfer
simulations to synthesize NIR scattered light images and mm dust
continuum maps. Spiral arms induced by GI, located at disk radii 
of hundreds of AUs, are local
overdensities and have their photospheres displaced
to higher altitudes above the disk
midplane; arms therefore scatter more NIR light from their central stars than inter-arm regions,
and are detectable at distances up to 1 kpc by Gemini/GPI, 
VLT/SPHERE, and Subaru/HiCIAO/SCExAO. By contrast, collapsed clumps formed
by disk fragmentation have such strong local gravitational fields that their scattering
photospheres are at lower altitudes; such fragments appear fainter
than their surroundings in NIR scattered light.
Spiral arms and streamers recently imaged in four FU Ori
systems at NIR wavelengths resemble GI-induced structures
and support the interpretation that FUors are
gravitationally unstable protostellar disks.
At mm wavelengths, both spirals and clumps appear brighter in thermal emission than
the ambient disk and can be detected
by ALMA at distances up to 0.4 kpc with one-hour integration
times at $\sim$0.1$\arcsec$ resolution. Collapsed
fragments having masses $\gtrsim 1 \mj$ can be detected by ALMA
within $\sim$10 minutes.

\end{abstract}



\section{Introduction}\label{sec:intro}

Both stellar and substellar companions can be spawned within gaseous accretion disks around newly born stars. At early times, the mass of the disk can be significant compared with the still-forming central star, and the mass infall rate from the disk's natal envelope can be larger than the accretion rate through the disk. Such young and unsteady disks experience gravitational instability (GI) if they satisfy the $Q$ criterion 
\citep{goldreich65}:
\begin{equation}
Q=\frac{c_{\rm s}\Omega}{\pi {\rm G} \Sigma}\lesssim {\rm unity},
\label{eq:q}
\end{equation}
where $c_{\rm s}$, $\Omega$, and $\Sigma$ are the disk's sound speed, orbital frequency, and surface density. Gravitationally unstable disks
have been simulated extensively \citep[e.g.,][]{vorobyov05, vorobyov06, vorobyov10burst, boley10, machida11, taukamoto13, zhu12planets}.
The disks may also fragment if they cool sufficiently fast \citep[e.g.,][]{gammie01, rice03, rafikov09, shi14}:
\begin{equation}
\beta=t_{\rm cool} \Omega \lesssim {\rm a\ few},
\label{eq:beta}
\end{equation}
where $t_{\rm cool}$ is the cooling timescale. Disk fragmentation can result in bound self-gravitating objects, and is thought to be a channel for the formation of giant planets, brown dwarfs, or stars.

We can enumerate a number of observational consequences of GI in protostellar/protoplanetary disks:
\begin{enumerate}
\item GI excites large-scale spiral arms, which may be visible in resolved observations at near-infrared \citep[NIR; e.g.,][]{dong15giarms, pohl15} and mm wavelengths \citep[e.g.,][]{cossins10,vorobyov13-observation, dipierro14}. 
\item At large orbital distances from the star, disk fragmentation may form bound objects such as giant planets, brown dwarfs, or stellar mass companions \citep[e.g., ][]{boss97, rice05, rafikov05, stamatellos09, kratter10, meru11, machida11, vorobyov10-planet, vorobyov13-planets}.
These objects, or their circumsecondary disks \citep[e.g.,][]{kraus15, caceres15}, may be detectable by adaptive optics (AO) imaging \citep[e.g.,][]{kraus12, bowler13, kuzuhara13}.
\item GI (possibly in combination with other disk instabilities such as the magnetorotational instability) can trigger outbursts \citep[e.g.,][]{vorobyov05, vorobyov06, vorobyov10burst, zhu09, machida11, bae14, vorobyov15}, detectable as abrupt and possibly repetitive surges in the accretion luminosity. The accretion outbursts in FU Orionis objects \citep[FUor;][]{hartmann98, audard14} and in some EX Orionis objects \citep[EXor;][]{herbig08} have been suggested to be triggered by GI.
\item The gas kinematics in self-gravitating disks deviates from Keplerian. These deviations may be detectable in molecular line observations \citep[e.g.,][]{rosenfeld14}.
\end{enumerate}

Historically, program 1 --- resolved imaging --- has been difficult to pursue.
Young disks are often embedded in and highly extincted by gaseous envelopes at optical to NIR wavelengths. FUors and EXors are rare and typically found in distant clusters of young stellar objects (e.g., Orion at $\sim$400 pc). To resolve GI-induced spiral arms and fragments at these distances demands 0.1$\arcsec$ or better angular resolution, and sub-arcsec inner working angles. Observations of possibly GI-unstable class 0/I objects (\citealt{andre94}) have so far been mostly indirect/photometric.

The situation, however, has been recently evolving, thanks to technological advances in both NIR direct imaging and mm-wave interferometry. At NIR wavelengths, a fleet of new instruments equipped with extreme adaptive optics and polarimetric differential imaging \citep[PDI; e.g.,][]{perrin04, hinkley09} has been deployed, including VLT/SPHERE \citep{beuzit08}, Gemini/GPI \citep{macintosh08}, and Subaru/HiCIAO/SCExAO \citep{jovanovic15}. At mm wavelengths, the new interferometers ALMA and JVLA are probing young stellar
objects with unprecedentedly high sensitivity and resolution. \citet{liu16fuori} directly imaged four FUors using Subaru/HiCIAO with sub-0.1$\arcsec$ angular resolution: FU Ori, V1735 Cyg, V1057 Cyg, and Z CMa (see also the VLT/NaCo observation of Z CMa by \citealt{canovas15zcma}). Polarization vector analysis confirmed that these objects contain disks. Large-scale asymmetric structures such as spirals and streamers, possibly produced by GI, were discerned at stellocentric distances of hundreds of AU. With its sub-0.05$\arcsec$ resolution, ALMA resolved the class 0/I disk HL Tau \citep{brogan15},
which has an estimated mass of $\gtrsim 0.1 M_\odot$ ($\gtrsim 0.2$ times the mass of its central star $M_\star$; \citealt{greaves08, kwon11}) and whose
self-gravity may be significant \citep{takahashi14, jin16}.

Looking ahead, we anticipate a great many images of class 0/I disks to become available at high angular resolution in the next decade. These new data will help to answer key questions in disk evolution and planet formation: How long is the GI-unstable phase? Do companions form through disk fragmentation? What are the nature of FUor and EXor outbursts? Motivated by these upcoming observations, in this paper we synthesize images of gravitationally unstable protostellar/protoplanetary disks at early times $\lesssim 0.5$ Myr, when 
infall from the envelope is still occurring. We employ the Vorobyov \& Basu (2015) hydrodynamics code to simulate the formation and evolution of a disk-star system directly from the collapse of a rotating, initially starless cloud core. The results of the hydro code are fed into the \citet{whitney13} Monte Carlo radiative transfer (MCRT) code to produce synthetic images of the disk. We focus on direct imaging of the NIR polarized intensity (PI), and on interferometric maps of the mm-wave dust thermal emission. The goal is to make predictions for future resolved observations, and in particular to answer the question of whether GI-induced spiral arms and fragments are detectable in NIR imaging observations using Gemini/GPI, VLT/SPHERE, and Subaru/HiCIAO/SCExAO, and in mm dust continuum observations using ALMA.

Some groundwork on synthesizing images has been laid. \citet{vorobyov13-observation} carried out semianalytic calculations to create mock ALMA observations of GI-unstable disks. They did not examine the properties of the disk in scattered light, nor did they consider non-face-on viewing angles. These limitations are removed in our work. \citet{stamatellos11} simulated ALMA observations of GI-unstable disks produced by their smoothed particle hydrodynamics code. We relax their assumptions that dust has a constant
temperature and is optically thin.
The appearance of GI-induced spiral arms and clumps has been studied by \citet{cossins10}, \citet{douglas13} and \citet{mayer16} at mm wavelengths and by \citet{dong15giarms} at NIR wavelengths; these works focused on isolated systems (i.e., those without
infalling envelopes) and on small scales ($\sim$100~AU); moreover,
the systems were only evolved for a few tens of thousand years or less.

The structure of our paper is as follows. We introduce our hydrodynamics simulation in Section~\ref{sec:hydro} and our MCRT simulations in Section~\ref{sec:mcrt}. The results of the hydro simulation are presented in Section~\ref{sec:results-hydro}, and synthetic images are shown in NIR scattered light in Section~\ref{sec:results-nir} and in the mm-wave thermal continuum in Section~\ref{sec:results-mm}. In Section~\ref{sec:limitations} we discuss the limitations of our models, and in Section~\ref{sec:prospect} we review the
near-term prospects for observing GI-unstable young disks. Section~\ref{sec:summary} summarizes.


\section{Simulations}\label{sec:setup}

We carry out a 2+1D hydrodynamics simulation to model the formation and evolution of a disk-star system directly from the collapse of an initially starless, rotating molecular cloud core. The hydro code (Section~\ref{sec:hydro}) is adopted largely from \citet{vorobyov15} with a few improvements as described below. The gas surface density is calculated from
the hydro simulation at 6 epochs. For
insertion into the \citet{whitney13} MCRT
code (Section~\ref{sec:mcrt}),
the 2D surface density from the hydro model is
``puffed up'' vertically into a 3D density field by
assuming hydrostatic equilibrium,
using an approximate temperature and vertical scale height 
inferred from the hydro model.
The MCRT code produces both scattered light 
and thermal emission maps; the latter
are computed from dust temperatures calculated in
detail from the MCRT model.

\subsection{Hydrodynamical Simulation}\label{sec:hydro}

We start our numerical simulation from the
gravitational collapse of a starless cloud core,
continue into the embedded phase of star formation ---
during which
a star, disk, and envelope are formed --- and terminate our
simulations at $t \sim 0.5$ Myr when the envelope
dissipates after accreting onto the star+disk system. 
Long integration times are made possible by the use
of the thin-disk (2D) approximation. This approximation enables us to follow the evolution of the disk for many orbital periods and its justification is discussed
in \citet{vorobyov10burst}. Once formed, the protostellar disk occupies the inner part 
of the numerical polar grid
(usually, several hundreds of AU), while the contracting envelope occupies the rest of the grid 
(which may extend to several thousands of AU).

To avoid time steps that are too short,
we introduce a ``sink cell'' at $r_{\rm sc}=10.0$~AU 
and impose a free outflow boundary condition so that 
matter is allowed to flow out of 
the computational domain but is prevented from flowing in. 
During the early stages of core collapse, we monitor the gas surface density in 
the sink cell; when its value exceeds a critical value for the transition from 
isothermal to adiabatic evolution, we introduce a first hydrostatic core (FHSC) with a size 
equal to that of the sink cell. The parameters of the FHSC are calculated assuming a simple 
polytropic sphere model with index $n = 2.5$ ($\gamma = 7/5$). 
When the temperature in the center of the FHSC exceeds 2000~K (triggering the dissociation of molecular hydrogen) or its mass exceeds 
0.05~$M_\odot$ (as suggested by radiation transfer calculations of \citealt{masunaga00}), we assume that the second collapse ensues and a central protostar forms.
In the subsequent evolution, 90\% of the gas that crosses the inner boundary 
is assumed to land on the protostar. 
The other 10\% of the accreted gas is assumed to be carried away by protostellar jets. 
The FHSC stage is usually short (tens of thousands of years) 
compared to the protostellar stage (several hundred thousand years).

The equations of mass, momentum, and energy transport in the thin-disk limit are
\begin{equation}
\label{cont}
\frac{{\partial \Sigma }}{{\partial t}} =  - \nabla_p  \cdot 
\left( \Sigma \bl{v}_p \right),  
\end{equation}
\begin{equation}
\label{mom}
\frac{\partial}{\partial t} \left( \Sigma \bl{v}_p \right) + \left[ \nabla \cdot \left( \Sigma \bl{v}_p
\otimes \bl{v}_p \right) \right]_p =   - \nabla_p {\cal P}  + \Sigma \, \bl{g}_p + (\nabla \cdot \mathbf{\Pi})_p
\end{equation}
\begin{equation}
\frac{\partial e}{\partial t} +\nabla_p \cdot \left( e \bl{v}_p \right) = -{\cal P} 
(\nabla_p \cdot \bl{v}_{p}) -\Lambda +\Gamma + 
\left(\nabla \bl{v}\right)_{pp^\prime}:\Pi_{pp^\prime}, 
\label{energ}
\end{equation}
where subscripts $p$ and $p^\prime$ refers to the planar components $(r,\phi)$ 
in polar coordinates, $\Sigma$ is the mass surface density, $e$ is the internal energy per 
surface area, 
${\cal P}$ is the vertically integrated gas pressure calculated via the ideal equation of state 
as ${\cal P}=(\gamma-1) e$,
$\bl{v}_{p}=v_r \hat{\bl r}+ v_\phi \hat{\bl \phi}$ is the velocity in the
disk plane, and $\nabla_p=\hat{\bl r} \partial / \partial r + \hat{\bl \phi} r^{-1} 
\partial / \partial \phi $. 
Turbulent viscosity is taken into account via the viscous stress tensor 
$\mathbf{\Pi}$, the expression for which is provided in \citet{vorobyov10burst}.
We parameterize the magnitude of kinematic viscosity $\nu$ using the $\alpha$-prescription 
with a spatially and temporally uniform $\alpha=0.01$.
The ratio of specific heats is calculated by assuming that $\gamma=5/3$ below 100~K
and $\gamma=7/5$ above 100~K. We apply a smooth transition at the critical temperature 
to avoid sharp changes in the values of $\gamma$. This form of $\gamma$ takes into account the
fact that the rotational and vibrational degrees of freedom of molecular hydrogen are excited
only above 100~K. 

The gravitational acceleration in the disk plane, $\bl{g}_{p}=g_r \hat{\bl r} +g_\phi \hat{\bl \phi}$, takes into account the self-gravity of the disk 
and the gravity of the central protostar when formed. The former component is found 
by solving for the Poisson integral
\begin{eqnarray} 
  \Phi(r,\phi) & = & - G \int_{r_{\rm sc}}^{r_{\rm out}} r^\prime dr^\prime 
                     \nonumber \\ 
      & &       \times \int_0^{2\pi} 
               \frac{\Sigma(r^\prime,\phi^\prime) d\phi^\prime} 
                    {\sqrt{{r^\prime}^2 + r^2 - 2 r r^\prime 
                       \cos(\phi^\prime - \phi) }}  \, ,
\end{eqnarray} 
where $r_{\rm out}$ is the radial position of the computational outer boundary (equivalently, the initial radius of the cloud core).
This integral is calculated using a FFT technique which applies the 2D Fourier 
convolution theorem for polar coordinates \citep[see][Sect.\ 2.8]{binney87}. 

An approximate expression for the
radially and azimuthally varying vertical scale height $h$
is determined in each computational cell via the equation of local vertical pressure balance \citep{vorobyov09}:
\begin{equation}
\rho \, \tilde{c}_s^2 = 2\int_0^h \rho \left( g_{z,\rm gas}+g_{z,\rm st} \right) dz,
\label{eq1}
\end{equation}
where $\rho$ is the gas volume density, $g_{z,\rm gas}$ and $g_{z,\rm st}$ are the {\rm vertical} 
gravitational accelerations due to disk self-gravity and the gravitational pull of 
the central star, respectively, and $\tilde{c}^2_{s}=\partial {\cal P} /\partial \Sigma$ is the 
effective sound speed of the non-isothermal gas. 
Assuming that $\rho$ is a slowly varying function of vertical distance $z$ between $z=0$ (midplane)
and $z=h$ --- i.e., $\Sigma=2\, h \,\rho$ --- and using Gauss's theorem, one can show that
\begin{eqnarray}
\int_0^h \rho \, g_{z,\rm gas} \,  dz &=& \frac{\pi}{4} G \Sigma^2\:, \label{eq2} \\
\int_0^h \rho \, g_{z,\rm st} \, dz &=& \frac{G M_\ast \rho}{r} 
\left\{  1-\left[ 1+ \left({\Sigma\over 2 \rho r} \right) \right]^{-1/2} \right\},
\label{eq3}
\end{eqnarray}
where $M_\ast$ is the mass of the central star.
Substituting equations~(\ref{eq2}) and (\ref{eq3}) back into equation~(\ref{eq1}), we obtain
\begin{equation}
\rho \, \tilde{c}_s^2 = {\pi \over 2} G \Sigma^2 + {2 G M_\ast \rho \over r} 
\left\{  1-\left[ 1+ \left({\Sigma\over 2 \rho r} \right) \right]^{-1/2} \right\}.
\label{height1}
\end{equation}
This can be solved for $\rho$ given the model's known 
$\tilde{c}_s^2$, $\Sigma$, and $M_\ast$, using Newton-Raphson iteration.
The vertical scale height is finally derived as 
\begin{equation}
h=\frac{\Sigma}{2\rho}.
\label{height2}
\end{equation}
This height is used to calculate
the stellar flux incident upon the disk surface, as per
equation (\ref{fluxF}) below, 
and to enable a connection
to our 3D MCRT code, as described in Section \ref{sec:mcrt}.

The radiative cooling per surface area in equation~(\ref{energ}) is determined using the diffusion
approximation for vertical radiation transport in a one-zone model (see the Appendix for details)
\begin{equation}
\Lambda=\frac{4\tau_{\rm P} \sigma \thydro^4 }{1+2\tau_{\rm P} + 
{3 \over 2}\tau_{\rm R}\tau_{\rm P}},
\end{equation}
where $\sigma$ is the Stefan-Boltzmann constant, 
$\thydro={\cal P} \mu / {\cal R} \Sigma$ is the temperature (``hydro'' to distinguish this temperature from the temperature
calculated from the MCRT code later),  $\mu=2.33$ the mean molecular weight, $\cal R$ the universal 
gas constant, $\tau_{\rm R}=\kappa_{\rm R} \Sigma_{1/2}$ 
and $\tau_{\rm P}=\kappa_{\rm P} \Sigma_{1/2}$ the 
Rosseland and Planck optical depths to the disk midplane,  and $\Sigma_{1/2}=\Sigma/2$ 
the gas surface density from the disk surface to the midplane.
The Planck and Rosseland mean opacities are calculated from the opacity tables by \citet{semenov03}.
The heating function per surface area of the disk is expressed as
\begin{equation}
\Gamma=\frac{4\tau_{\rm P} \sigma \tirr^4 }{1+2\tau_{\rm P} + {3 \over 2}\tau_{\rm R}\tau_{\rm
P}},
\end{equation}
where $\tirr$ is the irradiation temperature at the disk surface 
determined by the stellar and background blackbody radiation fields as
\begin{equation}
\tirr^4=T_{\rm bg}^4+\frac{F_{\rm irr}(r)}{\sigma},
\label{fluxCS}
\end{equation}
where $T_{\rm bg}$ is the uniform background temperature (set to the 
initial temperature of the natal cloud core)
and $F_{\rm irr}(r)$ is the radiation flux (energy per unit time per unit surface area) 
absorbed by the disk surface from the central star. The latter quantity is calculated as 
\begin{equation}
F_{\rm irr}(r)= \frac{L_\ast}{4\pi r^2} \cos{\gamma_{\rm irr}},
\label{fluxF}
\end{equation}
where $\gamma_{\rm irr}$ is the incidence angle of 
radiation arriving at the disk surface (with respect to the normal). The incidence
angle is calculated using the disk vertical scale height $h$, as described in \citet{vorobyov10burst};
disk self-shielding is not taken into account in the present study. 

The stellar luminosity $L_\ast$ is the sum of the accretion luminosity 
$L_{\rm \ast,accr}=(1-\epsilon) G M_\ast \dot{M}/2
R_\ast$ arising from the gravitational energy of accreted gas and
the photospheric luminosity $L_{\rm \ast,ph}$ due to gravitational contraction and deuterium burning
in the stellar interior. The stellar mass $M_\ast$ and accretion rate onto the star $\dot{M}$
are determined from the amount of gas passing into 
the sink cell. The properties of the forming protostar ($L_{\rm \ast,ph}$ and radius $R_\ast$) are 
calculated using the Lyon stellar evolution code described in \citet{baraffe10}.
As in \citet{baraffe12}, we assume that a fraction $\epsilon$ of the accretion energy  
$G M_\ast{\dot M}/(2 {R_\ast})$ is absorbed by the protostar, while the remaining fraction 
$(1-\epsilon$) is radiated away and contributes to the accretion luminosity of the star
$L_{\rm \ast,accr}$. 
Despite many efforts, the exact value of $\epsilon$ in low-mass star formation is not 
known. In the present calculations, we adopt a so-called ``hybrid'' scheme 
\citep[for details, see][]{baraffe12} with $\epsilon=0$ when accretion rates remain 
smaller than  a critical value $\dot{M}_{\rm cr}=10^{-5}~M_\odot$~yr$^{-1}$, and 
$\epsilon=0.2$  when $\dot{M} >  \dot{M}_{\rm cr}$.
The stellar evolution  code is coupled with the main hydrodynamical code in real time. 
Because of the heavy computational load, the stellar evolution code updates the properties of the protostar 
only every 20~yr, while the hydrodynamical time step may be as short as a few months.

Equations~(\ref{cont})--(\ref{energ}) are solved 
in polar coordinates on a numerical grid with $512 \times 512$ grid zones. 
The radial grid zones are logarithmically spaced, while the grid spacing in the 
azimuthal direction is uniform. The details of the solution procedure are given in
\citet{vorobyov10burst}. For initial conditions, we considered a gravitationally unstable 
pre-stellar core with the following radial profiles
of column density $\Sigma$ and angular velocity $\Omega$:
\begin{eqnarray}
\Sigma & = & {r_0 \Sigma_0 \over \sqrt{r^2+r_0^2}}\:, \\
\Omega & = &2\Omega_0 \left( {r_0\over r}\right)^2 \left[\sqrt{1+\left({r\over r_0}\right)^2
} -1\right],
\label{ic}
\end{eqnarray}
where $\Sigma_0=5.2\times10^{-2}$~g~cm$^{-2}$ and 
$\Omega_0=1.25$~km~s$^{-1}$~pc$^{-1}$ are the gas surface density and angular velocity at the center of the core. These profiles have a small near-uniform
central region of size $r_0=2400$~AU which transitions to an $r^{-1}$ profile;
they are representative of a wide class of observations and theoretical models
\citep{basu97,andre93,dapp09}. The initial radius of the core is 0.07~pc and its initial 
temperature is uniform at 10~K. The total mass of the core is 1.08~$M_\odot$.

\subsection{Monte Carlo Radiative Transfer Simulations}\label{sec:mcrt}

The MCRT simulations largely follow the procedures in \citet{dong15gaps}, and are summarized below. We construct a 3D grid in the radial ($r$), azimuthal ($\phi$), and polar ($\theta$) directions. The grid covers from the dust sublimation radius $\rsub$ to an outer radius $\rout=1000$ AU, $0-2\pi$ in $\phi$, and $0-\pi$ in $\theta$ ($\theta=\pi/2$ is the disk midplane), and has $419\times512\times200$ cells in $r\times\phi\times\theta$. At $r\geq\rin$ the $r\times\phi$ grid is identical to the polar grid in the hydro model, while from $\rsub\leq r<\rin$ (a region not covered by the hydro model) the grid is logarithmic in $r$. The sublimation radius $\rsub$ is determined for each epoch of the hydro model as where the dust temperature reaches 1600 K ($\rsub\sim0.1$~AU). The grid spacing $\delta\theta$ is linearly proportional to $\theta$ to better resolve the disk midplane. At $1000\geq r({\rm AU})\geq20$, $\sigmag$ is set by the hydro model; at $20\,{\rm AU}\geq r\geq\rsub$, $\sigmag$ follows a $1/r$ radial profile. We note that the envelope at $r >1000$ AU in the hydro simulation is not included in the MCRT calculations; this omission will be discussed in Section~\ref{sec:prospect}. To construct 3D disk models from 2D hydro $\sigmag$ maps, we assume hydrostatic equilibrium in the vertical direction $z$, and puff up the 2D disk according to a Gaussian profile,
\begin{equation}
\rhog(z)=\frac{\sigmag}{h\sqrt{2 \pi}} e^{-z^2/2h^2}, \label{eq:gaussian}
\end{equation}
where $\rhog$ is the gas volume density and the scale height $h$ is computed from the hydro simulation (Equation~\ref{height2}). This treatment is necessary as 2D hydro simulations are needed to follow the evolution of the system for hundreds of thousands of years, while 3D disk structures are required in the radiative transfer post processing. Millimeter fluxes from our models are not affected by this treatment. At NIR wavelengths, contrasts of disk features may be weakened \citep{zhu15densitywaves}, which enhances our conclusion that GI-induced spiral arms are visible (see below). The central star in MCRT simulations has the same temperature, mass, and luminosity (photospheric+accretion) as determined in the hydro simulation at each epoch (Table~\ref{tab:models}). MCRT simulations have only one illumination source (the central star), i.e., the fragments are not self-luminous. All simulations are run with at least 2 billion photon packets.

Millimeter observations have shown evidence of grain growth in class 0/I objects. For example, the $\sim$1~Myr old HL Tau disk is very bright at mm wavelengths, indicating the presence of large grains \citep{brogan15}. Also, \citet{miotello14} and \citet{chiang12} found low values of the mm spectral index in class I objects. To take grain growth into account, we include two populations of dust grains in the MCRT simulations: small and big grains. Small grains are standard interstellar medium (ISM) grains \citep{kim94} made of silicate, graphite, and amorphous carbon. Their size distribution obeys a power law in the size range of $0.02\lesssim s\lesssim0.25$~$\micron$, followed by an exponential cut off at larger sizes. Big grains have identical composition, but have grown to a maximum size of 1~mm with a power law size distribution
$dn(s)/ds\propto s^{-3}$.
The optical properties of both populations can be found in Figure~2 of \citet{dong12cavity}. The opacity of the big grains
is 13 cm$^2$ g$^{-1}$ at 1.3 mm (dust only, not dust+gas). The scattering phase function of the small grains is approximated using the \citet{henyey41} function, and a Rayleigh-like phase function is assumed for the linear polarization \citep{white79}.

Dust grains dominate the opacity in disks. Specifically, NIR scattered light arises from the disk surface, and mainly probes the distribution of small grains (both their surface density and vertical distribution). Mm continuum emission is sensitive to the surface density distribution of big grains. To convert $\rhog$ into $\rhod$, we assume a total dust-to-gas mass density ratio of 1:100 and a small-to-big dust mass density ratio of 1:9. We assign $\rhosg=10^{-3}\rhog$ (small grains are well-mixed with gas). For the big grains, we take $\Sigma_{\rm big\ grains}=9\times10^{-3}\sigmag$ and distribute them vertically according to a Gaussian with scale height $h_{\rm big\ grains}=0.5h$ to mimic vertical settling. We note that the specific ratio of $h_{\rm big\ grains}/h$ is not important as long as it is less than 1; the 
NIR scattered light is more sensitive to the small grains, and the mm continuum emission is not sensitive to the vertical distribution of the big grains.

Full resolution synthesized PI images at $\lambda= 1.6 \,\micron$ ($H$-band) and mm continuum maps at 1.3 mm (230 GHz; ALMA band 6) and 0.87 mm (345 GHz; ALMA band 7) are produced from the MCRT simulations.\footnote{In this work, the physical quantity recorded in all synthetic images is the specific intensity in units of
[mJy~arcsec$^{-2}$] ([$10^{-26}$ ergs~s$^{-1}$~cm$^{-2}$~Hz$^{-1}$~arcsec$^{-2}$]), or [mJy~beam$^{-1}$].} Full resolution $H$-band images are then convolved by a Gaussian PSF with a full width half maximum (FWHM) of 0.04$\arcsec$, to simulate the diffraction limited angular resolution of Subaru, VLT, and Gemini. Full resolution mm images are transformed to simulated ALMA observations using the {\tt simobserve} and {\tt simanalyze} tools under Common Astronomy Software Applications (CASA). A full array of 50 12-meter antennas is used.\footnote{The configurations of the full array are listed at https://casaguides.nrao.edu/index.php?title=Antenna \_Configurations\_Models\_in\_CASA. The angular resolution decreases with increasing configuration number.} Throughout the paper, we use a blue-hot color scheme at $H$-band and a red-hot color scheme for ALMA images. We assume a source distance of 400 pc (the distance to the Orion star forming region) unless noted otherwise.


\section{Results}\label{sec:results}

In this section, we examine the outcomes from both the hydro and MCRT simulations. The general evolution of the disk, the behavior of episodic accretion, and the formation and properties of fragments in the hydro simulations have been explored extensively in the series of papers by Vorobyov et al. In this work we focus on the visibility of GI-induced fragments and spiral arms in resolved images. To reiterate some of the global model parameters from Section~\ref{sec:hydro}, the initial radius of the core is 0.07 pc, its total mass is $1.08\msun$, and the ratio of its  initial rotational energy to its gravitational potential energy is 0.68\%.

\subsection{Hydro Models}\label{sec:results-hydro}

\begin{figure}
\begin{center}
\includegraphics[trim=0 0 0 0, clip,width=\textwidth,angle=0]{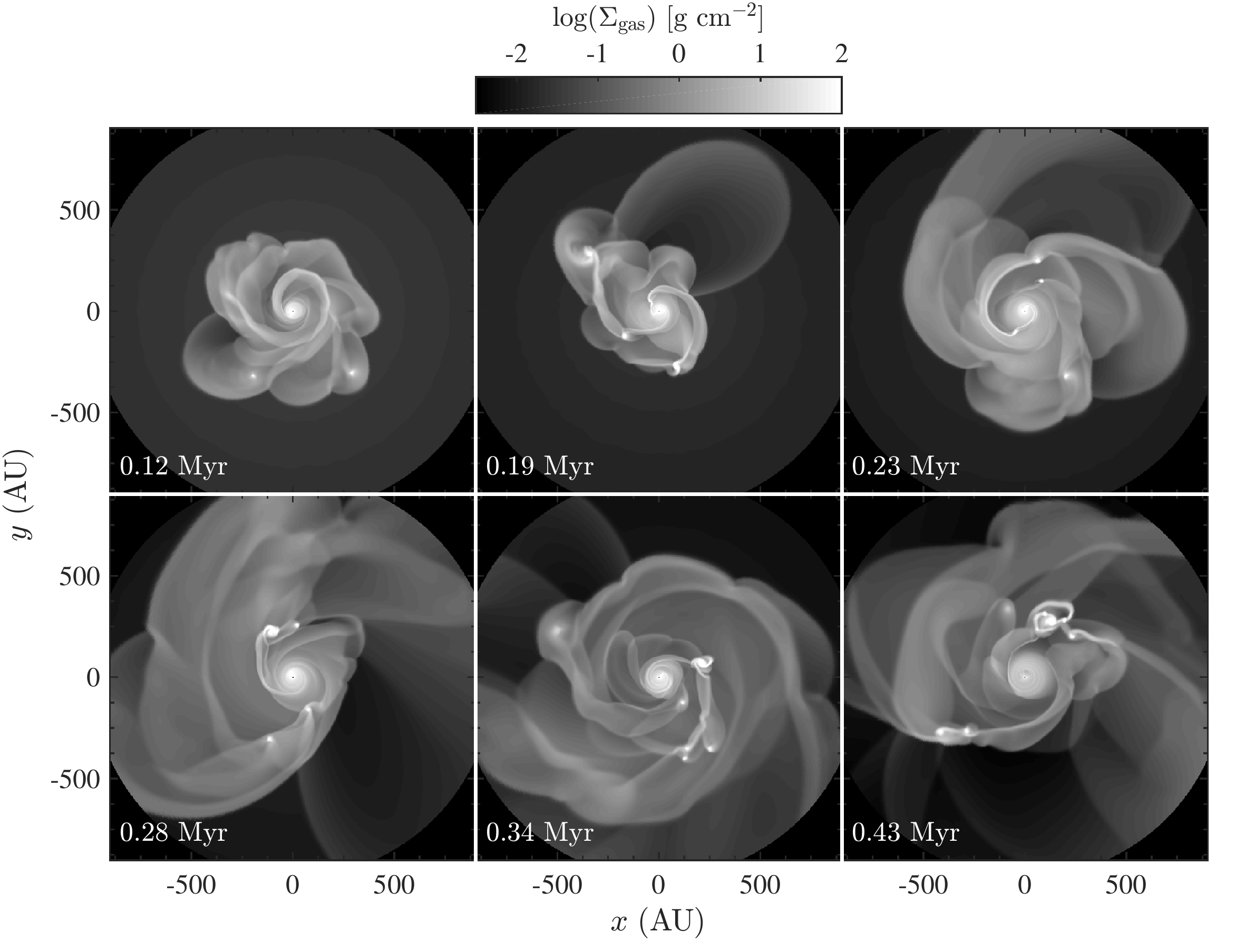}
\end{center}
\figcaption{Model gas surface densities. The disk rotation is counterclockwise.
\label{fig:sigma}}
\end{figure}

Figure~\ref{fig:sigma} shows $\sigmag$ at $t=0.12$, 0.19, 0.23, 0.28, 0.34, and 0.43 Myr in the hydro simulation. Time here is counted from the formation of the protostar (and not from the onset of molecular cloud collapse). A rotating disk component emerges around the protostar at $\sim$0.016~Myr. The disk mass grows from $0.13 \msun$ at t=0.12 Myr to $0.18 \msun$ at t=0.32~Myr; at the meantime the stellar mass increases from $0.37 \msun$ to $0.61 \msun$ (the disk-to-star mass ratio gradually declines with time). Soon after the disk forms, it undergoes GI as driven by ongoing mass loading from the infalling envelope, and fragments. The disk radius increases from $\sim$300~AU at $\sim$0.1 Myr to a maximum of $\sim$700 AU at $\sim$0.3 Myr, and afterwards shrinks. At the 6 epochs, 3--7 fragments are present at stellocentric distances between 100--600 AU. The fragments are generally smaller than 40 AU (0.1$\arcsec$ at 400~pc), and are characterized by surface densities of $\sim$50--2000 g cm$^{-2}$. Spiral arms and lobes are present at all epochs. These structures are located at hundreds of AUs from their central stars, and can often extend over $\pi$ in the azimuthal direction. The surface density $\sigmag$ of the arms is usually 2.5--20$\times$ higher than the azimuthal average of the background disk at the same radius.

\begin{figure}
\begin{center}
\includegraphics[trim=0 0 0 0, clip,width=\textwidth,angle=0]{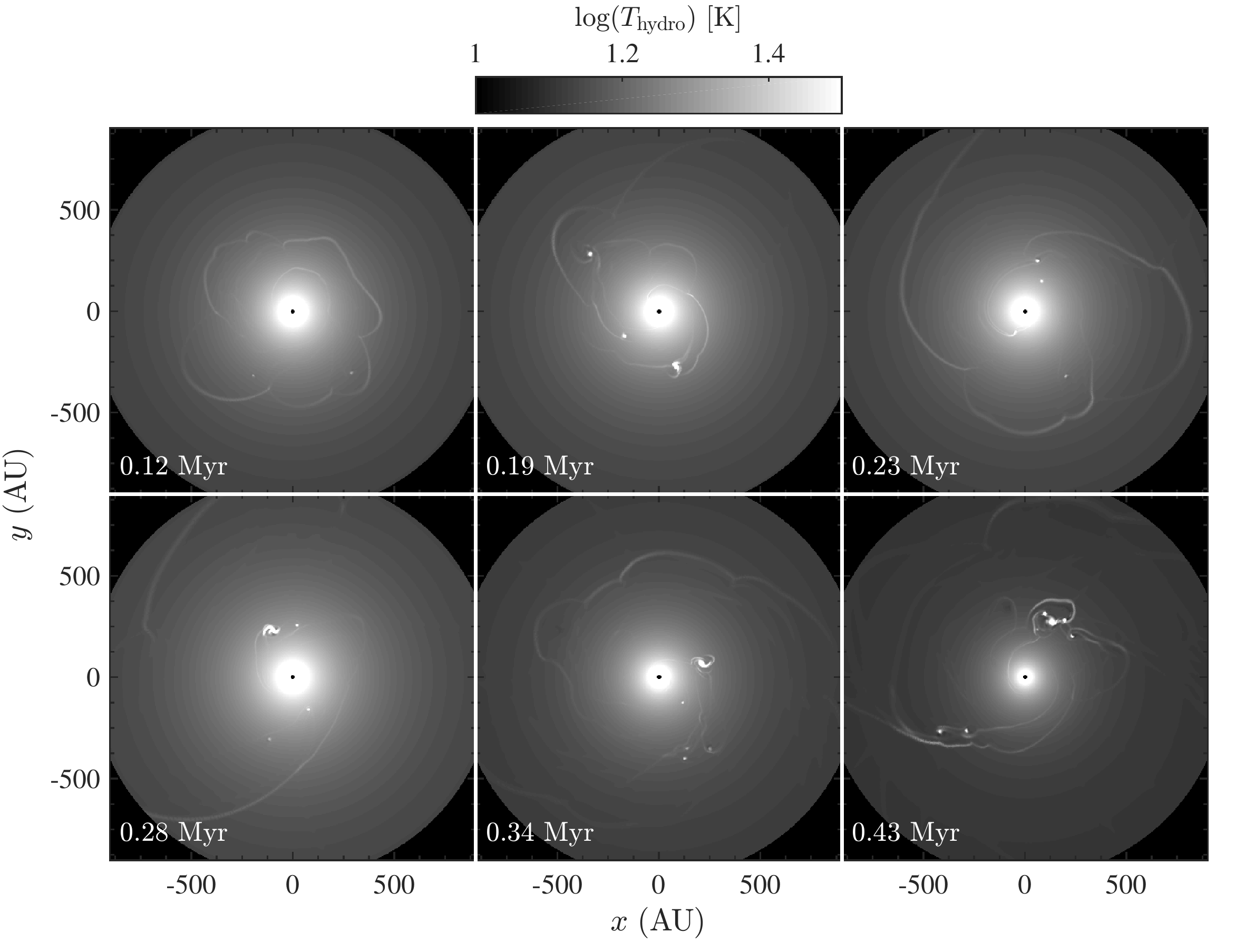}
\end{center}
\figcaption{Temperature in the hydro models $\thydro$.
\label{fig:thydro}}
\end{figure}

\begin{figure}
\begin{center}
\includegraphics[trim=0 0 0 0, clip,width=\textwidth,angle=0]{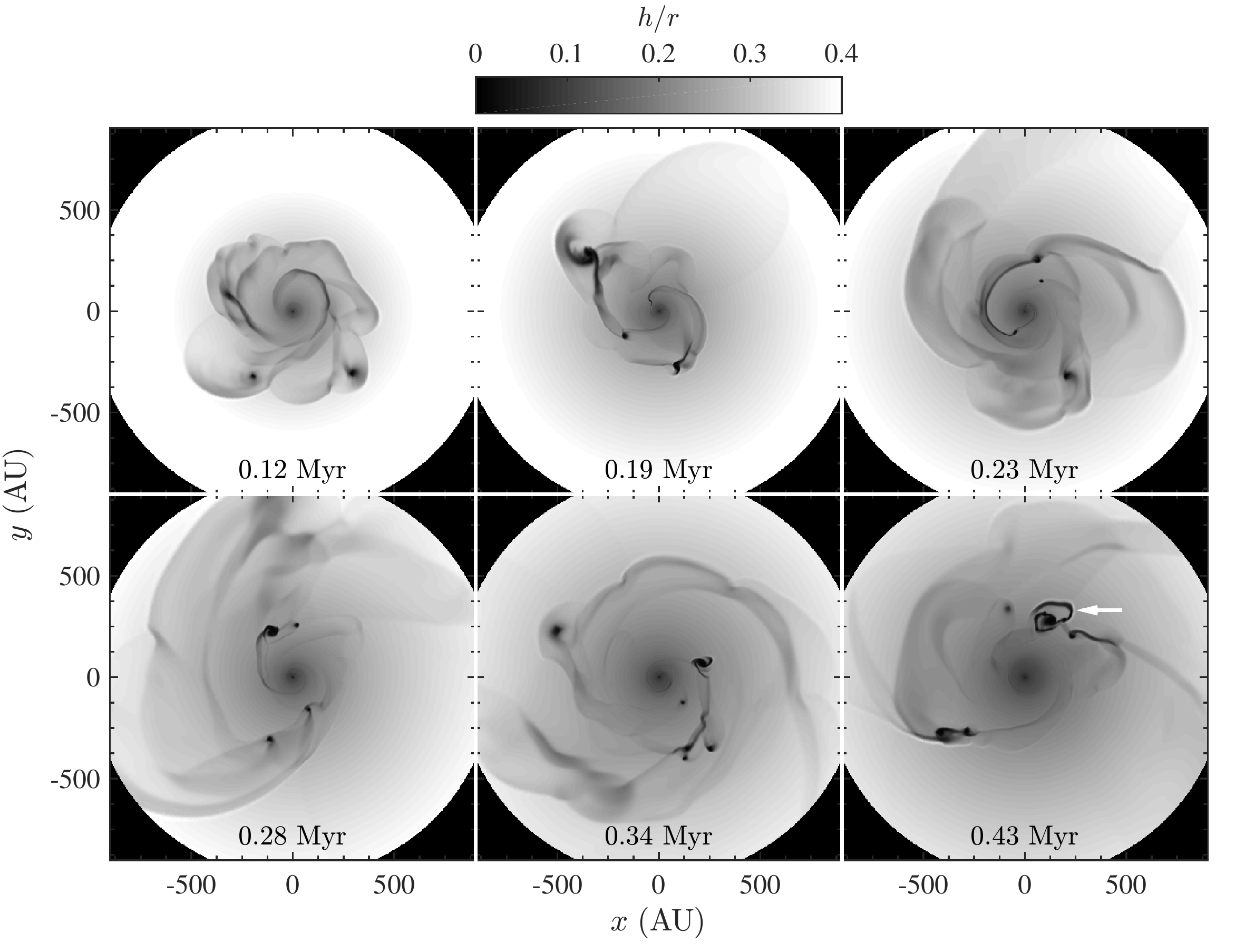}
\end{center}
\figcaption{Disk aspect ratios $h/r$ in the hydro models.
\label{fig:hoverr}}
\end{figure}

Figures \ref{fig:thydro} and \ref{fig:hoverr} show $\thydro$ and $h/r$ from the hydro simulation. As derived in Section~\ref{sec:hydro}, $h$ follows from gravity (from both the central star and disk) balancing pressure $P$ in the vertical direction, where $P$ is set by the hydro disk temperature $\thydro$. Azimuthally averaged $\thydro$ vary from $\sim 80$ K at 10 AU to $\sim 30$  K  at 100 AU. In high density regions such as spiral arms and fragments, $\thydro$ increases to up to 230~k because $PdV$ work
is done to compress the gas, increasing $h$.
On the other hand, in these same overdensities, 
the local gravity is stronger, decreasing $h$.
Our results show that the latter effect dominates the former so that $h$ decreases in high density regions. In most spiral arms,  $h$ drops by $\lesssim20\%$. Such a drop does not necessarily
imply that the NIR scattering surface is lower inside arms
than outside arms; we will find in Section \ref{sec:results-nir}
that the scattering surface is actually higher inside the arms
because they contain more material. In fragments,
the collapse of the local
scale height is more dramatic: $h$ drops by factors of 3--30.

\subsection{NIR Images}\label{sec:results-nir}
 
\begin{figure}
\begin{center}
\includegraphics[trim=0 0 0 0, clip,width=0.8\textwidth,angle=0]{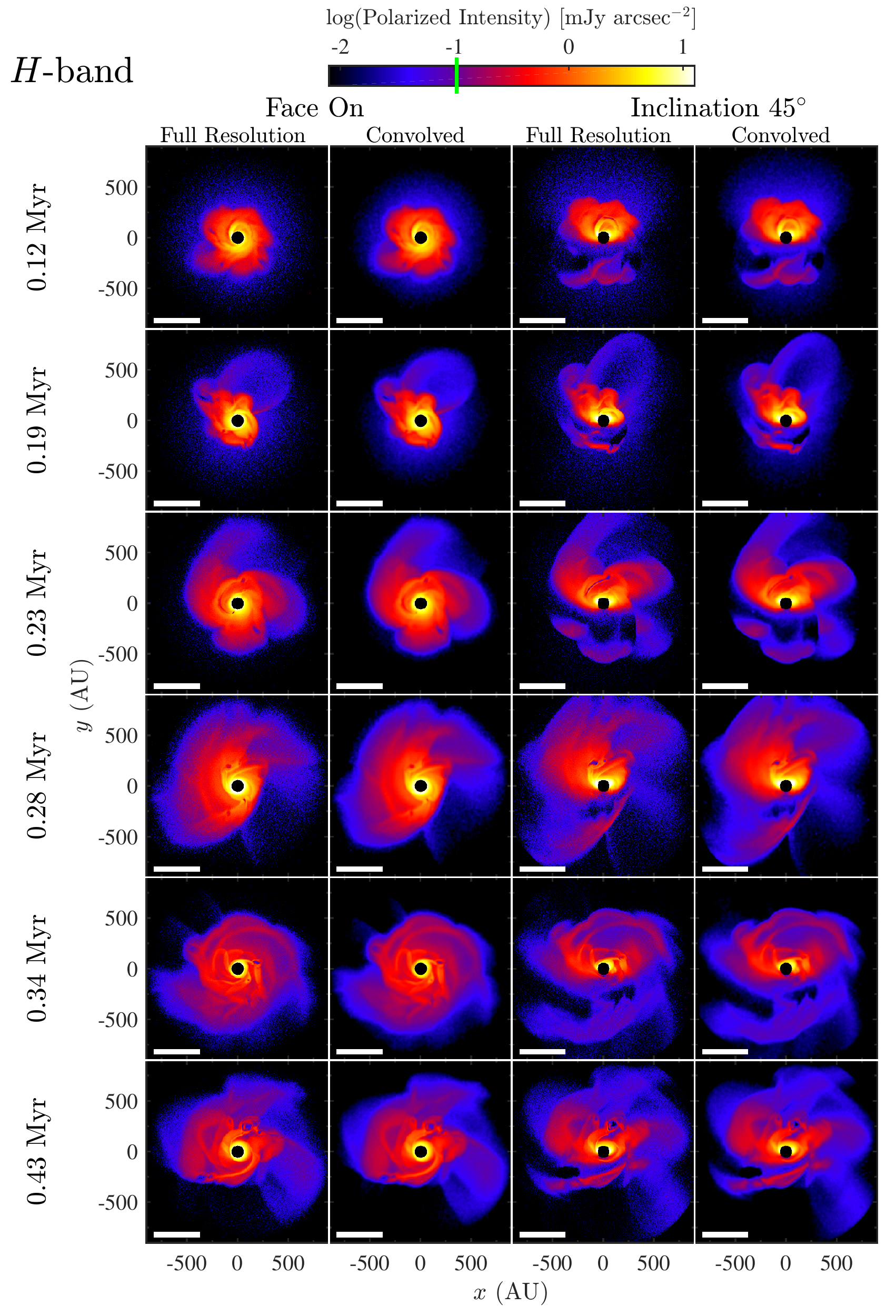}
\end{center}
\figcaption{\footnotesize $H$-band PI images of the disk viewed face-on (left two columns) and at an inclination of $45^\circ$ (right two columns; the side of the disk nearest the observer is located at the bottom of each panel, and the major axis is horizontal). The $1^{\rm st}$ and $3^{\rm rd}$ columns are at full resolution, while the $2^{\rm nd}$ and $4^{\rm th}$ columns contain images convolved with a Gaussian PSF with a FWHM of 0.04$\arcsec$ (angular resolution of an 8-m telescope at $H$-band) at a source distance of 400 pc. Current detection limits (noise levels) for NIR PI imaging are $\sim$0.1 mJy arcsec$^{-2}$, corresponding to the transition between red (detected) and blue (undetected). Each panel masks out an inner working angle of  $0.15\arcsec$. The scale bar at the lower left is $1\arcsec$ long. See Section~\ref{sec:results-nir} for details.
\label{fig:image_h}}
\end{figure}

\begin{figure}
\begin{center}
\includegraphics[trim=0 0 0 0, clip,width=0.7\textwidth,angle=0]{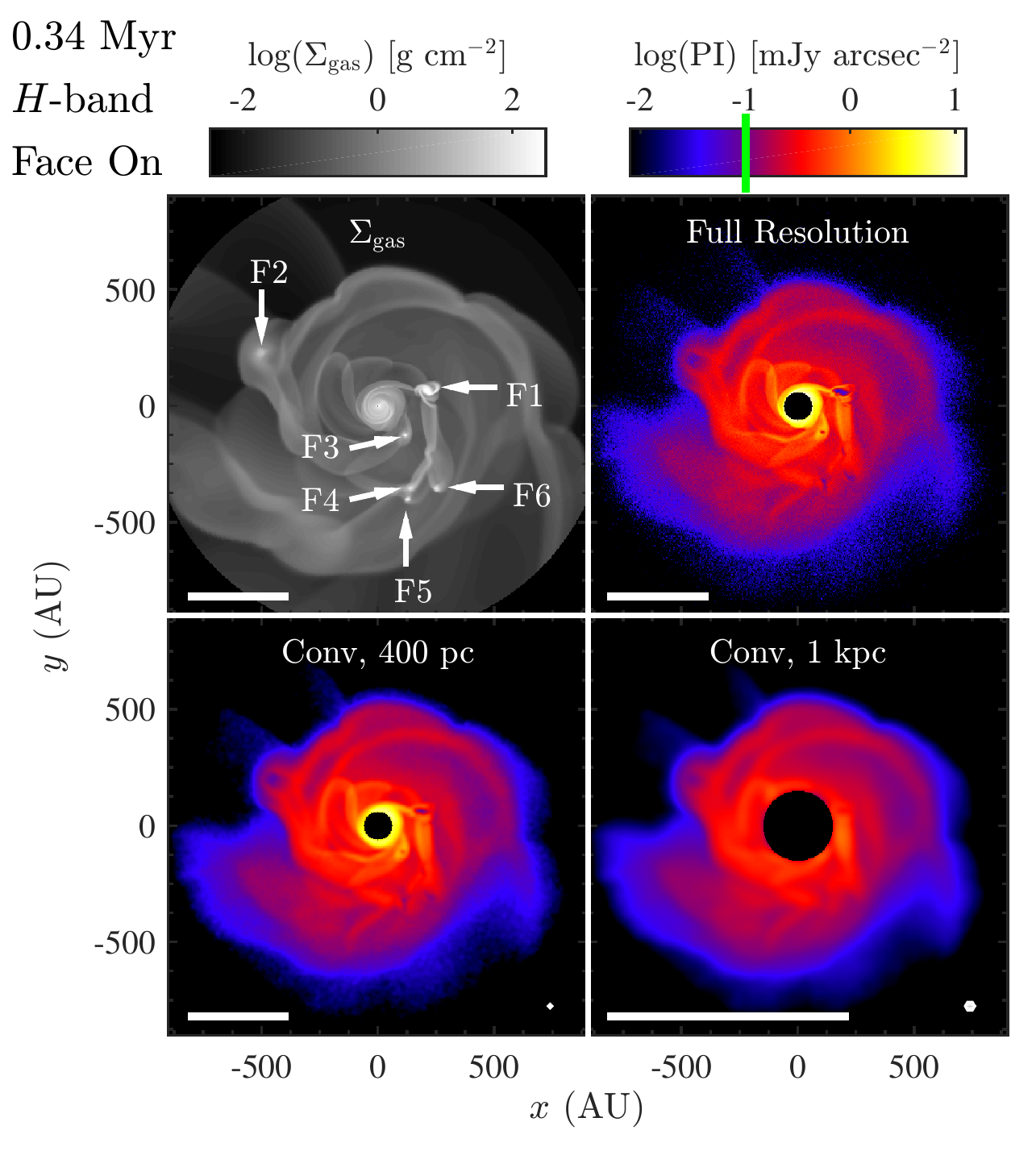}
\end{center}
\figcaption{Comparing gas surface density to NIR surface brightness at 0.34 Myr, and the effect of source distance. The bottom two panels are convolved images assuming a distance of 400 pc (left) and 1 kpc (right). The locations of 6 fragments are marked in the $\sigmag$ map. The central $0.15\arcsec$ in each NIR image is masked out, and the scale bar is $1\arcsec$. Current detection limits (noise levels) for NIR PI imaging are $\sim$0.1 mJy arcsec$^{-2}$, marked as the green tick on the color bar. See Section~\ref{sec:results-nir} for details.
\label{fig:image_h_comp}}
\end{figure}

Full resolution and convolved $H$-band PI images of disks at a distance of 400 pc are shown in Figure~\ref{fig:image_h} for all 6 epochs and 2 viewing angles (face-on and $45^\circ$ inclination). A detailed comparison between these images at $t=0.34$ Myr and the raw $\Sigma$ image is made in Figure~\ref{fig:image_h_comp}. At all epochs the disk is bright and shows complicated structures. We note that the current detection limit (noise level) in NIR PI imaging observations lies at about 0.1 mJy arcsec$^{-2}$, if
not lower \citep[e.g.,][]{hashimoto12, mayama12, kusakabe12, grady13, follette13}.\footnote{0.1 mJy arcsec$^{-2}$ is the detection limit for AO188+HiCIAO onboard Subaru. Detection limits for the newer generation of instruments, such as Gemini/GPI and VLT/SPHERE, are expected to be better.} This corresponds to the transition between blue (undetectable) and red (detectable) in our NIR color scheme. Our assumed angular resolution ($0.04\arcsec$) is small enough to resolve most spiral arms. As a result, the convolved images are quite similar to the full resolution images.

Most spiral arms appear as prominent bright features, while all fragments appear as depressions in surface brightness (see the 6 fragments marked in Figure~\ref{fig:image_h_comp}). These brightness variations are caused by variations in the height of the scattering photosphere, defined as the surface where the optical depth $\tau_\star$ to the star is 1. This $\tau_\star=1$ surface is determined by $\Sigma$ and $h$. As shown in \citet{juhasz15}, a change in $h$ of at least $20\%$ is required for a structure to be visible in current NIR observations. In our simulated spiral arms, the drop in $h$ due to self-gravity is generally $\lesssim 20\%$ and therefore insignificant;\footnote{There are a few, extremely dense spiral arms for which the depression in $h$ is severe; an example is indicated by the arrow in Figure~\ref{fig:hoverr}.}
thus, the increase in surface density within the arms pushes the
local $\tau_\star = 1$ surface higher than the surrounding background. Spirals arms are therefore illuminated by the star
and appear brighter. By contrast, in fragments, the drop in $h$ due to self-gravity is so significant that these regions are shadowed and appear as holes in NIR scattered light.

Figure~\ref{fig:image_h_comp} also demonstrates the effect of increasing the distance to the object to 1~kpc (roughly the distance to FUors V1735 Cyg and Z CMa). The major spirals at $r\gtrsim200$~AU remain visible as their physical sizes (in particular their widths) are comparable to the resolution.

The spiral arms appear similar when viewed face-on or at $45^\circ$ inclination (see the right two columns of Figure~\ref{fig:image_h}). The edge of the nearside of the disk (the bottom side) is not always parallel to the (horizontal) major axis because of variations in disk surface density and scale height. At 0.28 Myr the disk is so asymmetric that even when viewed face-on, it appears nearly one-sided.

\subsection{Millimeter Images}\label{sec:results-mm}

\begin{figure}
\begin{center}
\includegraphics[trim=0 0 0 0, clip,width=0.8\textwidth,angle=0]{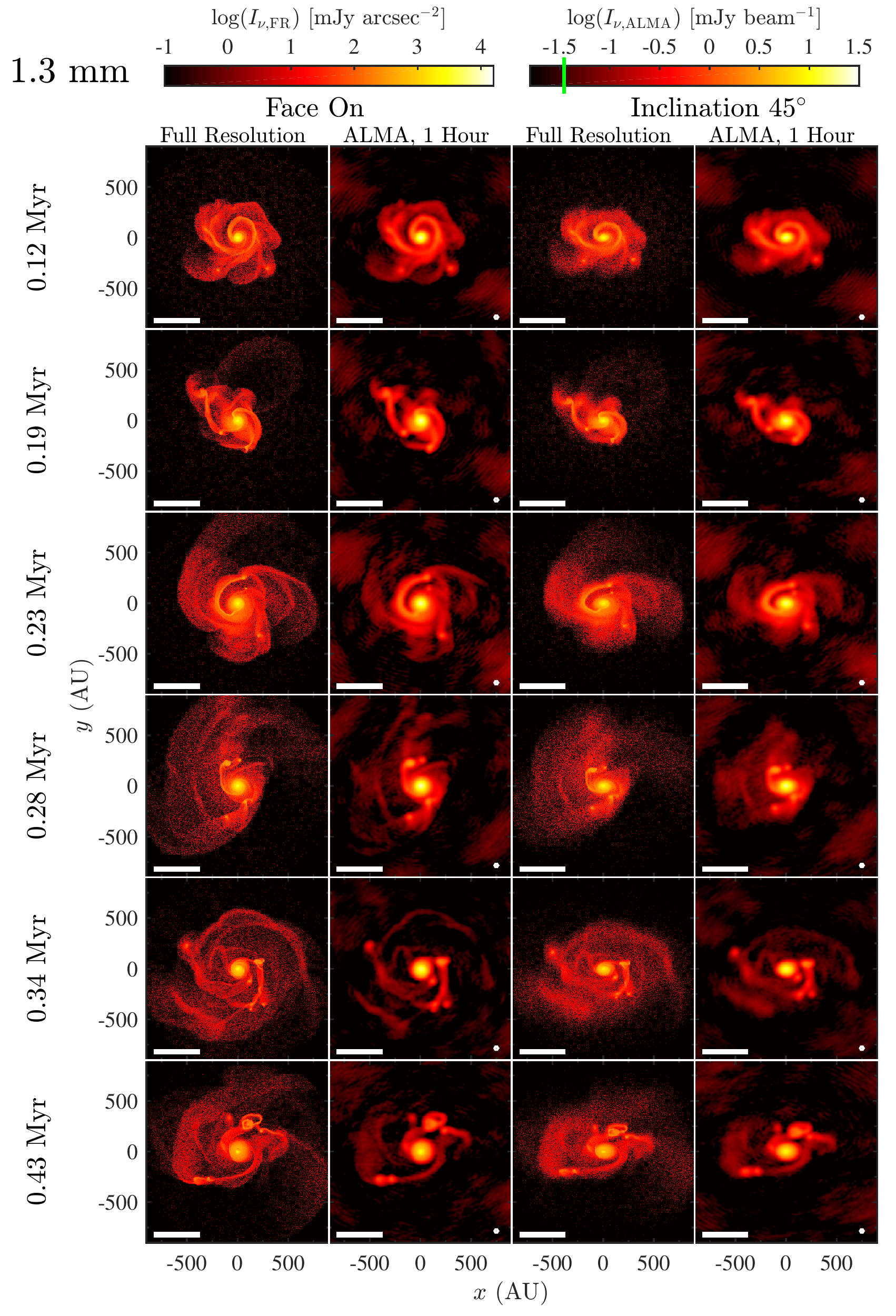}
\end{center}
\figcaption{\footnotesize ALMA band 6 (1.3~mm) images at viewing angles of $0^\circ$ (left two columns) and $45^\circ$ (right two columns; the side of the disk nearest the observer is oriented toward the bottom of each panel, and the major axis is horizontal). The $1^{\rm st}$ and $3^{\rm rd}$ columns are at full resolution and use the color bar on the left. The $2^{\rm nd}$ and $4^{\rm th}$ columns, using the color bar on the right, are simulated one-hour integrations with ALMA in array configuration \#19 (angular resolution $\sim$0.1$\arcsec$; the beam is indicated in the lower right corner). The source distance is 400 pc. The $3\sigma$ detection limit in simulated ALMA images is 0.034 mJy beam$^{-1}$ and is marked by the green tick on the color bar. The horizontal scale bar indicates $1\arcsec$. See Section~\ref{sec:results-mm} for details.
\label{fig:image_z1}}
\end{figure}

\begin{figure}
\begin{center}
\includegraphics[trim=0 0 0 0, clip,width=\textwidth,angle=0]{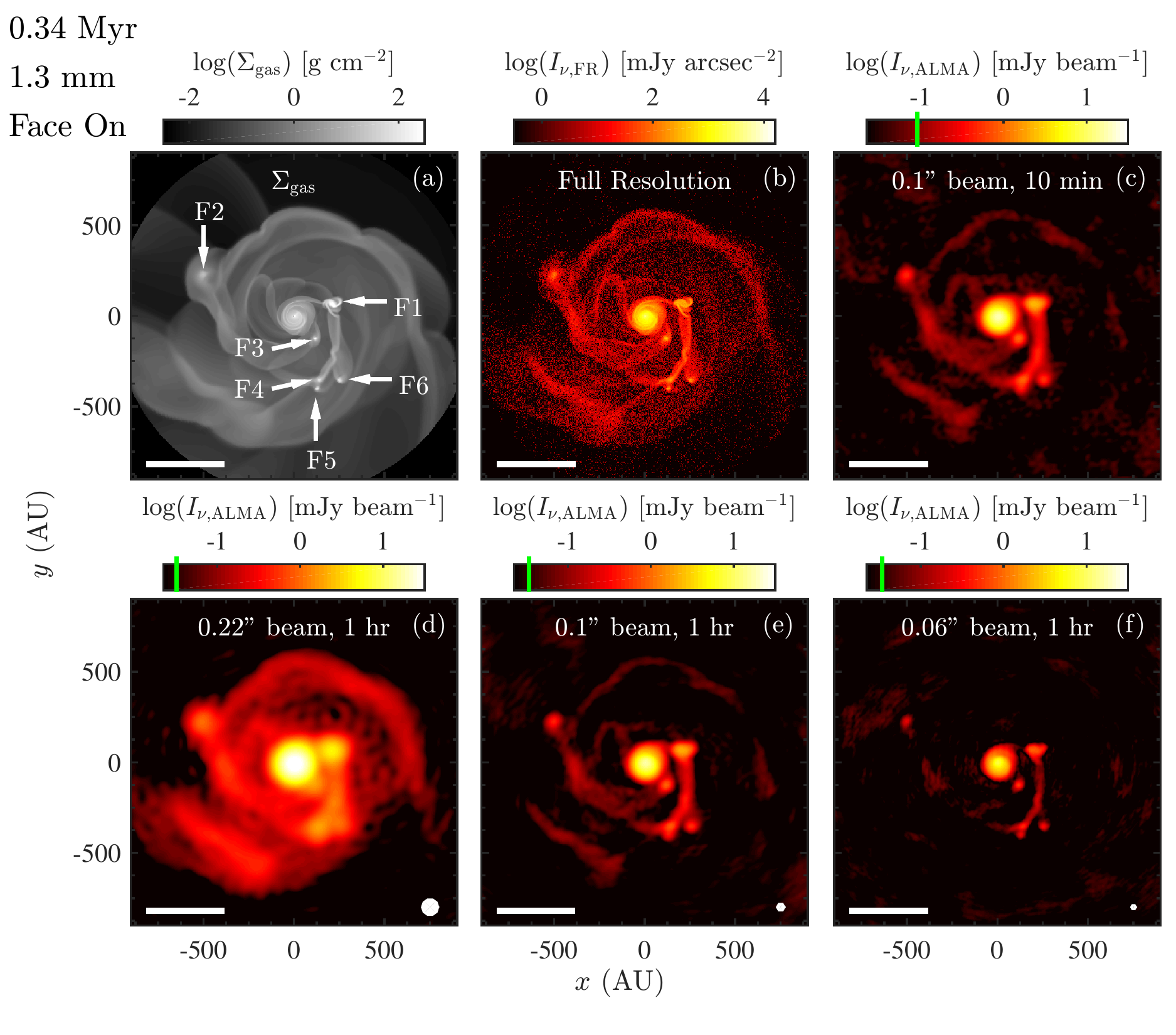}
\end{center}
\figcaption{The effects of integration time and angular resolution on ALMA observations. Panels (c), (d), (e), and (f) are simulated ALMA observations based on the full resolution synthetic MCRT image in (b). The source is at 400 pc. The array configurations used in the ALMA simulator are \#13 (d; angular resolution $\sim$$0.22\arcsec$), \#19 (b, e; angular resolution $\sim$$0.1\arcsec$), and \#22 (d; angular resolution $\sim$$0.06\arcsec$). The difference between (c) and (e) is integration time: 10 minutes vs.~1 hour. The $3\sigma$ detection limit in each simulated ALMA image is marked as a green tick on the color scale. See Section~\ref{sec:results-mm} for details.
\label{fig:image_z1_comp}}
\end{figure}

Figure~\ref{fig:image_z1} shows full resolution MCRT dust continuum images and simulated ALMA images at 1.3 mm (230 GHz; ALMA band 6). Full resolution images closely trace the surface density at all 6 epochs.
Figure~\ref{fig:image_z1_comp} provides a closer look at 0.34 Myr, at which time all 6 fragments and major spiral arms identified
in $\Sigma$ (panel a) are clearly visible in the full resolution image (panel b). 
By contrast to the NIR, both spiral arms and fragments in the thermal continuum appear as local maxima, despite the fact that the surface temperatures of the fragments are lower than their surroundings because they are shadowed. Fragments appear bright in the mm
because they are optically thick, whereas the ambient disk
is optically thin, with vertical optical depths
$\lesssim 0.05$.

\begin{figure}
\begin{center}
\includegraphics[trim=0 0 0 0, clip,width=0.85\textwidth,angle=0]{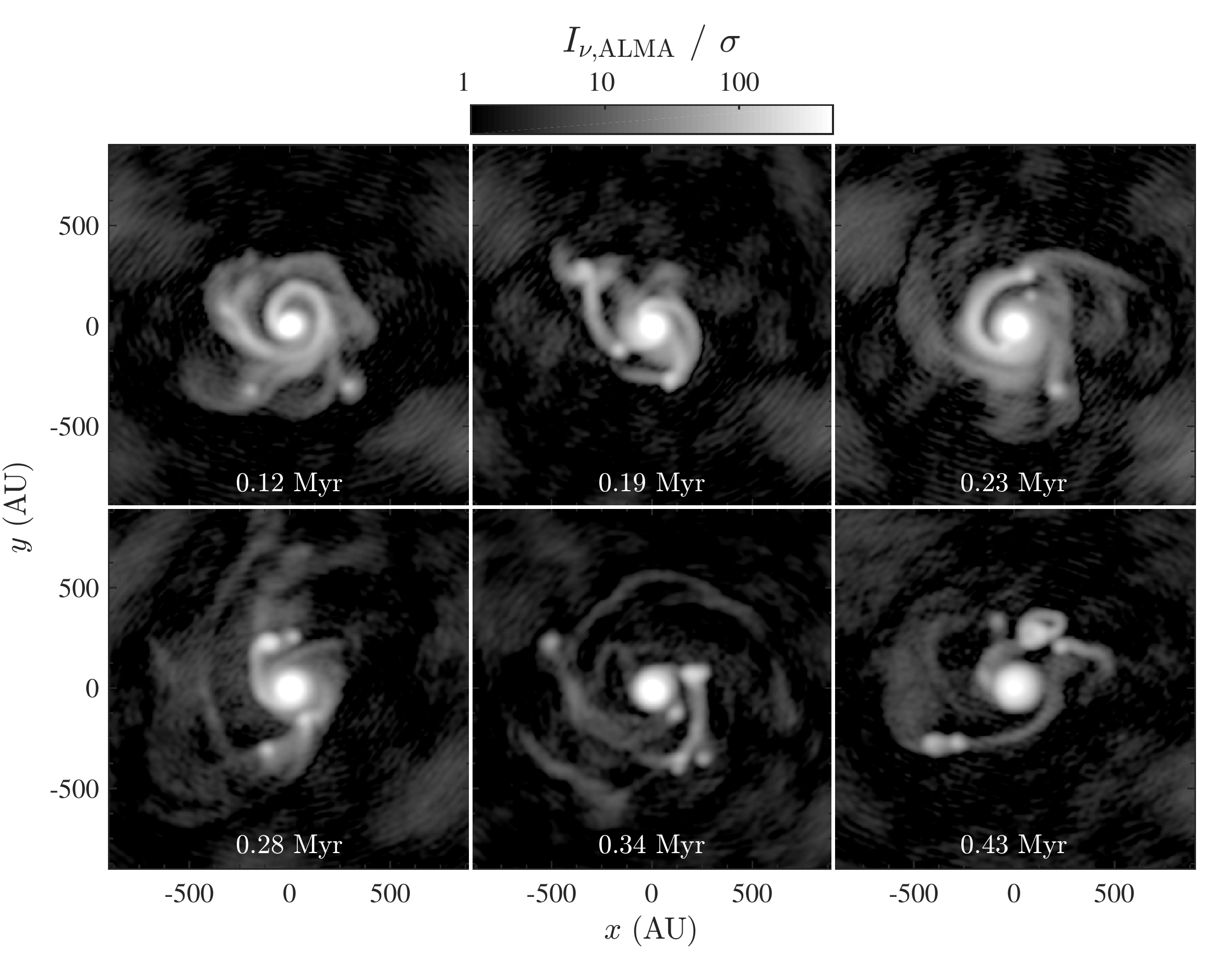}
\end{center}
\figcaption{The statistical significance of detections in 
simulated face-on ALMA Band 6 observations (see also Figure~\ref{fig:image_z1}), in units of $\sigma=0.011$~mJy beam$^{-1}$. See Section~\ref{sec:results-mm} for details.
\label{fig:image_z1_significance}}
\end{figure}

The simulated ALMA images in Figure~\ref{fig:image_z1} are produced with array configuration \#19 and integration times of 1 hour. The 3$\sigma$ detection limit is 34 $\mu$Jy beam$^{-1}$ as calculated by the ALMA sensitivity calculator;\footnote{https://almascience.eso.org/proposing/sensitivity-calculator. Default parameters for water vapor column density, $T_{\rm sky}$, and $T_{\rm sys}$ are adopted.}
in our color scheme this noise floor corresponds to the transition between black (undetected) and red (detected). The synthetic beam size is $\sim$0.1$\arcsec$ (40 AU at 400 pc).\footnote{Briggs weighting with the default {\tt robust=}0.5 is used in {\tt simanalyze} to achieve a compromise between minimizing side lobes and minimizing the noise level.} Evidently, $0.1\arcsec$ angular resolution is sufficient to resolve most fragments and spiral arms at 400 pc. This can be further illustrated by comparing panels (a) and (e) in Figure~\ref{fig:image_z1_comp}. Among the 6 fragments in (a), F1, F2, F3, and F6 are clearly visible and distinguishable in (e), while F4 and F5 may be difficult to separate because of their small separation ($\sim$50~AU). The sensitivity is sufficient to detect emission from all major structures out to hundreds of AUs. Figure~\ref{fig:image_z1_significance} shows the significance of detections for face-on ALMA images in Figure~\ref{fig:image_z1}. The masses of fragments, determined using the fragment tracking algorithm of \citet{vorobyov13-planets}, are 33, 1.0, 2.7, 2.7, 1.9, and 2.5 $\mj$, respectively (note that F2 is likely still in the process of formation and will continue growing in the subsequent evolution). All fragments are significantly detected in (c) by at least $40\sigma$, even for the lowest mass fragment F2 of 1$\mj$ (also the lowest mass fragment among all epochs). Spiral arms are generally detected at $\gtrsim10\sigma$ except in the outer regions beyond $\sim$700~AU. Simulated ALMA observations using the parameters underlying Figure~\ref{fig:image_z1} recover $>70\%$ of the total mm flux density, as listed in Table~\ref{tab:models}. These conclusions are consistent with those of \citet{vorobyov13-observation}.

The angular resolution and the total integration time are two key parameters in ALMA observations. Figure~\ref{fig:image_z1_comp} illustrates the effects of varying these two parameters. Panels (c) and (e) demonstrate the difference between an integration time of 10 minutes (c) and 60 minutes (e), while the resolution is fixed at $0.1\arcsec$. Since the sensitivity is inversely proportional to $\sqrt{\rm integration\ time}$, the noise level in (c) is 2.4 times higher than in (e). The {\it uv} sampling of the 10-minute snapshot observation is less complete, and thus side lobes are more prominent in (c). Nevertheless, qualitatively the two panels are quite similar to each other. Fragments F1, F2, F3, F4+F5, and F6 are all significantly detected in (c), with the the least massive fragment F2 (1 $\mj$) detected at $17\sigma$. Most spiral arms are also visible in (c), though some are only marginally detected.

Panels (d), (e), and (f) demonstrate the effect of different angular resolutions, achieved by varying the array configuration: (d) has an angular resolution of $0.22\arcsec$ (88 AU at 400 pc, or 2.2 times larger than (e)), while (f) has an angular resolution of $0.06\arcsec$ (24 AU at 400 pc, or $40\%$ smaller than (e)).\footnote{ALMA array configurations \#13 and \#21 are used for (d) and (f), respectively.} The integration time is fixed (1 hour) so that their sensitivities are the same in units of mJy beam$^{-1}$. The impact of angular resolution is dramatic. In (d), F4, F5, and F6 merge into a larger clump, and F3 is absorbed into the central peak. Spiral arms that are close together also merge to form larger (wider) arms. In (f), although all 6 fragments are resolved by the small beam, their detection significance drops: the weakest source F2 is now a $10\sigma$ detection. At small angular resolution, large-scale structures, such as most spiral arms, are severely resolved out and not visible. As a result, (f) only recovers about $40\%$ of the total flux density in the original full resolution image, while (d) recovers $90\%$ and (e) recovers $72\%$ of the total flux density.

\begin{figure}
\begin{center}
\includegraphics[trim=0 0 0 0, clip,width=0.8\textwidth,angle=0]{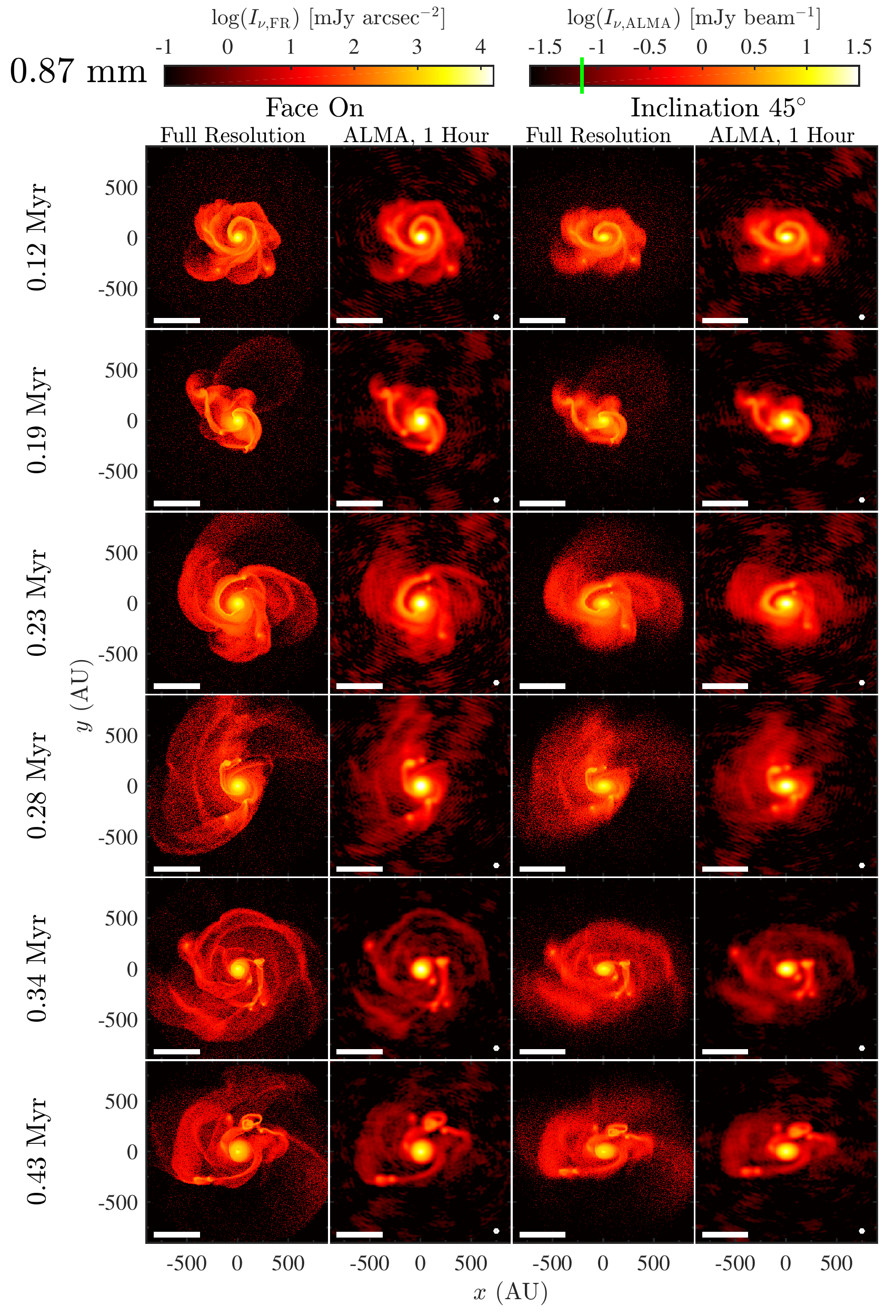}
\end{center}
\figcaption{Same as Figure~\ref{fig:image_z1}, but for ALMA Band 7 (0.87 mm), array configuration \#16, and a one-hour integration time.
\label{fig:image_a7}}
\end{figure}

Synthetic ALMA observations at 0.87~mm (345 GHz; ALMA band 7) with $0.1\arcsec$ angular resolution (produced with array configuration \#16) are shown in Figure~\ref{fig:image_a7}. Qualitatively they are similar to the 1.3~mm images.


\section{Discussion}\label{sec:discussions}

\subsection{Limitations of Our Models}\label{sec:limitations}

\subsubsection{Gas-Dust Decoupling}
Grains in disks are subject to both gravity and gas drag \citep[e.g.,][]{weidenschilling77, birnstiel10}. Particles with 
 dimensionless stopping times (a.k.a. Stokes numbers)
\begin{equation}
\tau_{\rm s}=\frac{\pi s \rho_{\rm bulk}}{2\sigmag}
\label{eq:ts}
\end{equation}
approaching unity drift the most quickly toward
gas pressure maxima
 \citep[e.g.,][]{rice06, zhu12}.
Here $\rho_{\rm bulk} \sim 1$ g/cm$^3$ is the internal
bulk density of a grain, and $s$ is the grain size.
In our models, big grains
with sizes $s \sim 1$ mm at $r\gtrsim300$~AU can have
$\tau_{\rm s} \sim 1$ in low-density inter-arm regions. As a result, our models may be overestimating the surface density of the big grains in low density regions at large distances, as big grains may actually be accumulating in high pressure regions such as spiral arms and fragments. Consequently, we may be underestimating the contrast of the arms and fragments at $r\gtrsim300$~AU in mm images relative to the surrounding background. Nevertheless, fragments and spiral arms are already detected with high significance in our simulated ALMA images, even without this further concentration; properly taking this effect into account can only enhance their visibilities.

\subsubsection{$T_{\rm hydro}$ vs.~$T_{\rm MCRT}$}
There are two versions of disk temperature in our models. The temperature from our hydro code, $\thydro$, takes into account both radiation from the central star and hydrodynamical processes such as $PdV$ work, viscous and shock heating, and is computed assuming the radiation field is diffusive (Appendix). On the other hand, the MCRT simulation calculates its own temperature, $\tmcrt$, which takes into account scattering, absorption, and re-emission of starlight in 3D, but does not include hydrodynamical processes. We expect $\tmcrt$ to be more accurate than $\thydro$, everywhere except in regions where hydrodynamical processes are significant, such as inside fragments. 

\begin{figure}
\begin{center}
\includegraphics[trim=0 0 0 0, clip,width=0.5\textwidth,angle=0]{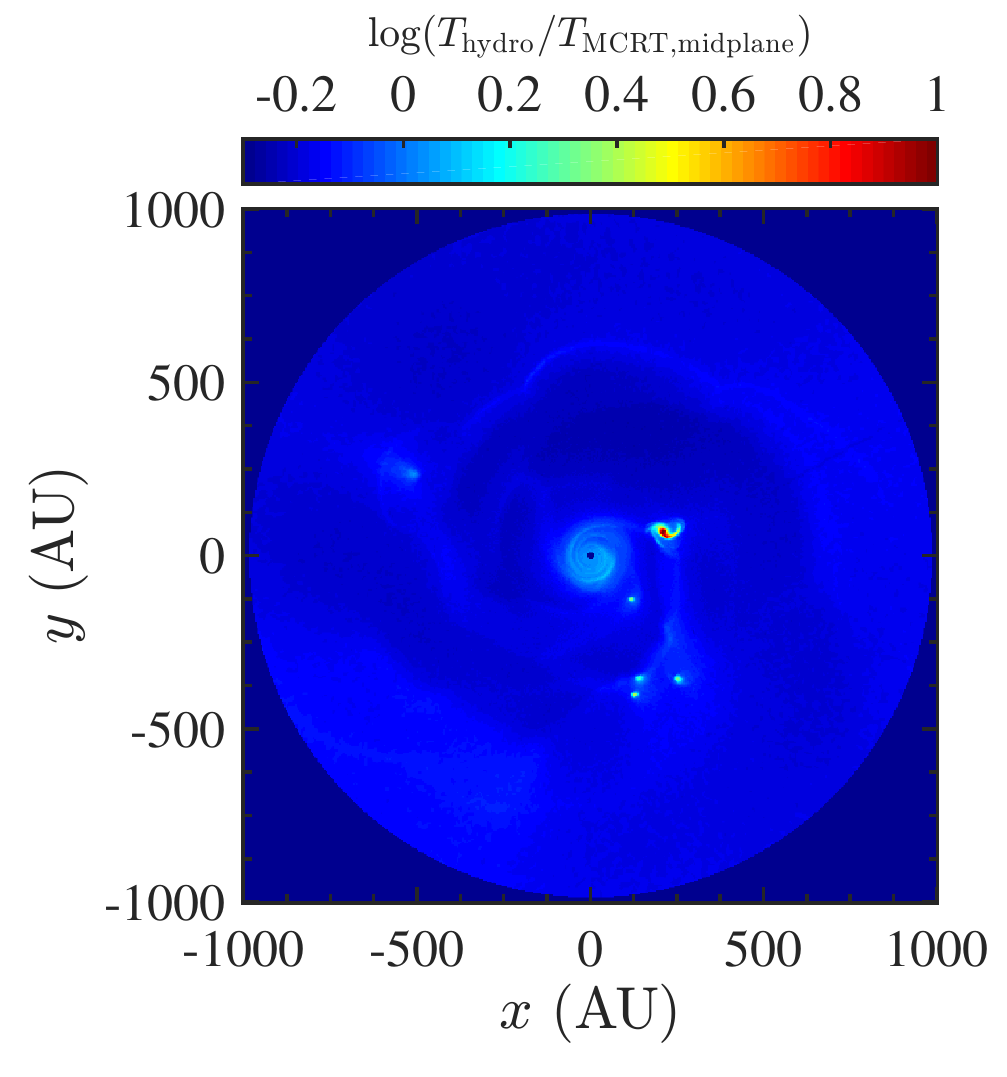}
\end{center}
\figcaption{Log$({\thydro/\tmcrtmid})$ at 0.34 Myr. The difference between the two temperatures is $\lesssim 40$\% everywhere except at the centers of fragments, where $\thydro$ can be up to one order of magnitude higher than $\tmcrtmid$. See Section~\ref{sec:limitations} for details.
\label{fig:t_comparison}}
\end{figure}

Figure~\ref{fig:t_comparison} compares the midplane temperatures
calculated by the two methods at 0.34 Myr. For the most part, except in fragments,
$\thydro$ is lower than $\tmcrt$ by $\lesssim40\%$.
Since $h\propto \sqrt{T}$, the inaccuracy in $h$ as propagated from $\thydro$ is expected to be $\lesssim20\%$. This is
not a major source of error; \citet{juhasz15} found that abrupt changes in $h$ of $\gtrsim 20\%$ are needed to generate discernible effects in current NIR direct imaging observations.

More serious is the discrepancy between temperatures calculated
within dense fragments.  In fragment centers, $\thydro$ can 
exceed $\tmcrt$ by up to one order of magnitude 
(Figure~\ref{fig:t_comparison}).
The mm-wave thermal fluxes of fragments may therefore be
underestimated by our MCRT simulations (the NIR images are more
reliable insofar as they depend on vertical scale heights 
calculated from the more realistic hydro simulation).  To assess 
the error in the 
mm-wave images, we re-calculate the mm fluxes by inserting the
$\thydro$ data into the 3D ray-tracing module of the {\tt NATALY}
radiative transfer code described in \citet{pavlyuchenkov11}.  In
these ray-tracing calculations, the disk temperature is set by
$\thydro$ and is vertically uniform; the disk is puffed up in the
vertical direction in the same way as for the MCRT simulations
(see equation \ref{eq:gaussian} and related text); and the (big) 
dust grains are assumed well mixed with gas with a dust-to-gas 
mass ratio of $0.9:100$ and an opacity identical to that of big 
dust in the MCRT
simulations.  To simulate the inner disk in these ray-tracing
calculations, we fill the inner 30~AU\footnote{The radius of the
ray-traced inner disk is larger than the sink cell radius 
$r_{\rm in} = 20$ AU used in the hydro simulations to ensure a 
smooth gas surface density profile.} with H$_2$ (number density 
$10^{10}$ cm$^{-3}$ and temperature 50 K).

\begin{figure}
\vspace*{-0.5cm}
\begin{center}
\includegraphics[trim=0 0 0 0, clip,width=0.7\textwidth,angle=0]{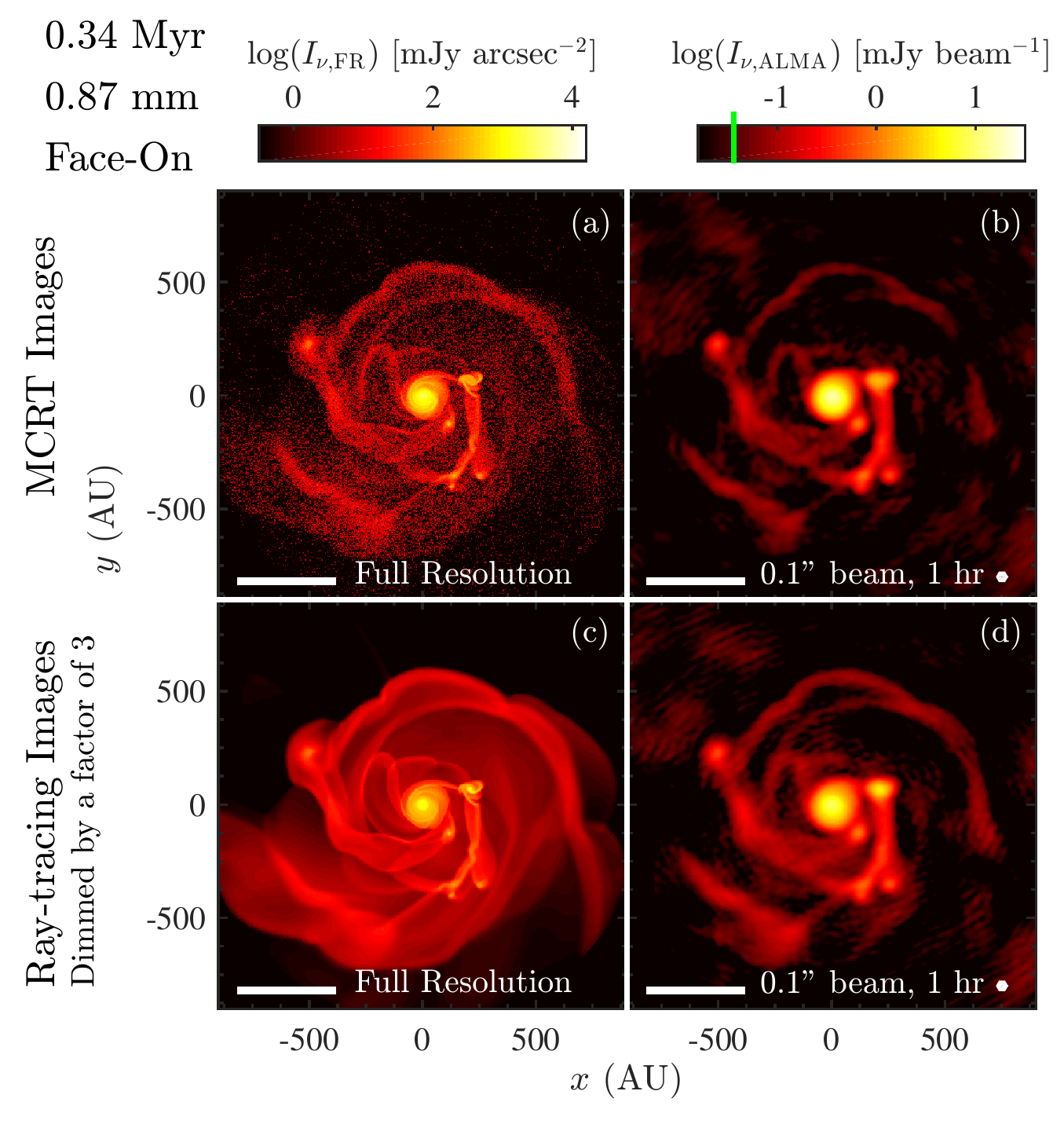}
\vspace*{-0.5cm}
\end{center}
\figcaption{Comparison between MCRT mm-wave images (top row) and mm-wave images produced by ray-tracing calculations with
{\tt NATALY} (bottom row). The source is at 400 pc distance. The left panels show full resolution images, while the right panels show simulated one-hour ALMA observations with array configuration \#19 (resolution $\sim0.1\arcsec$). The ray-traced images are dimmed by a factor of 3 to fit within the color scheme adopted for the MCRT images. See Section~\ref{sec:limitations} for details.
\label{fig:image_z1_yaroslav}}
\end{figure}

Figure~\ref{fig:image_z1_yaroslav} compares the ray-traced
mm images with the MCRT images at 0.34 Myr.
The ray-traced images are dimmed by a factor of 3 to fit within the same color scheme used for the MCRT images. Qualitatively the morphology of the disk in the two sets of images are similar, with the peak MCRT fluxes lower by a factor of $\sim$3.
This factor of 3 difference in flux
is less than the factor of 10 difference in midplane
temperature because the fragments are optically thick.
In the end, the MCRT and ray-traced images agree that GI-induced spirals and fragments in class 0/I disks at 400 pc can be detected and resolved by ALMA with 1-hour integration times at $0.1\arcsec$ angular resolution. 

\subsection{Near-Term Prospects for Observing GI in Disks}\label{sec:prospect}

Disks with $Q<1$ fragment on dynamical, i.e., orbital timescales. Fragments produced by GI may appear quickly, within $\sim$10$^4$ years,
reducing the disk mass and stabilizing the system against
further activity \citep[e.g.,][]{stamatellos11}. Thus GI
may be a short-lived phenomenon that is difficult to observe.
However, GI can be recurring and prolonged if fresh gas is supplied to the disk from its natal envelope, on timescales up
to a few $\times$ 10$^5$ yr. Similarly, individual fragments may be lost (or downsized) as they get shredded by tidal forces, migrate into the central star, or get ejected from the system \citep[e.g.,][]{vorobyov10-planet, boley10, nayakshin10, machida11, zhu12planets, basu12, taukamoto15}; but fragments can be recurring as well.

At mm wavelengths, envelopes do not much obscure our view
of embedded disks. Contamination from envelope emission
is on the order of $\sim$30\% in dust continuum emissions for class 0 sources,
and $\sim$10\% for class I sources \citep{jorgensen09}.
At NIR wavelengths, the situation is more challenging,
as disks can be heavily extincted by envelopes.
Our MCRT calculations ignore envelopes and should therefore
be applied to systems in kind (i.e., late stage I), or
to disks with envelopes viewed nearly face-on with large opening
angles for their bipolar cavities.


We may search for GI in
(1) disks with $\mdisk\gtrsim0.3\mstar$, and (2) disks undergoing accretion outbursts. Millimeter observations of class 0/I disks have confirmed the presence of large ($>100$~AU) disks in a few systems \citep{tobin15, yen15, choi07}, and have suggested a number of candidates with $\mdisk \gtrsim 0.1M_\odot$ \citep{jorgensen07, jorgensen09, eisner08}, modulo the usual uncertainties in gas-to-dust ratio and dust opacity \citep[e.g.,][]{dunham14}. A particularly interesting case is HL Tau \citep{brogan15}, which has an estimated disk mass of about $0.1M_\odot$, or $20\%$ of the stellar mass \citep{greaves08, kwon11}. \citet{jin16} have suggested that the disk is marginally GI-unstable, and that disk self-gravity facilitates the formation of gaps by relatively low-mass planets \citep[e.g.,][]{dong15gaps, dipierro15-hltau}. 

Young stellar objects undergoing accretion outbursts, such as FUors and EXors, are thought to be GI-unstable disks. Measuring their disk masses \citep[e.g.,][]{liu16} will be crucial for validating this interpretation. These objects are excellent targets for future high angular resolution observations to detect GI-induced spiral arms. In pioneering work, \citet{liu16fuori} directly imaged spiral arms and streams in four FUors with Subaru. The structures seen in the Subaru images resemble those in our NIR model images, supporting the idea that FUors are GI-unstable protostellar/protoplanetary disks  \citep[e.g.,][]{vorobyov05, vorobyov10burst, vorobyov15}.

What are the implications if the clumps and spiral arms
predicted by our synthetic images are not seen?
\begin{enumerate}
\item If mm continuum observations do not detect the disk while NIR imaging observations do, it may imply that we have overestimated the mm-wave dust opacity. In our model, the mm opacity of the big grains (up to 1 mm in size) is about 13 cm$^2$ g$^{-1}$, consistent with the results found by \citet{draine06} to within factors of a few for grains of similar sizes, and about two orders of magnitude higher than the corresponding opacity for ISM grains. If substantial grain growth has not occurred in class 0/I disks, the mm fluxes of our models could be overestimated by up to two orders of magnitude.
\item If mm observations reveal no Keplerian disk beyond a few tens of AU (e.g., as suggested in B335 by \citealt{yen15nodisk}), it may imply the action of strong magnetic braking \citep{krasnopolsky02, li11, machida11magneticbraking}, enabled perhaps by very small grains $\sim$10--100 $\AA$ in size that can couple magnetic fields to matter \citep{zhao16}. NIR imaging observations may not see these small disks at all if they lie inside the inner working angle (a few tens of AU at a few hundred pc).
\item If observations at both NIR and mm wavelengths show a large but featureless (i.e., axisymmetric) disk on scales $\gtrsim 100$ AU, the disk
may either be insufficiently massive to be GI-unstable
(equation \ref{eq:q} is not satisfied) or cool on timescales
so long that GI-induced activity is too anemic to be detected
($t_{\rm cool} \Omega \gg 1$; cf.~equation \ref{eq:beta}).
\end{enumerate}


\section{Summary}\label{sec:summary}

Using 2+1D hydrodynamics simulations, we modeled the formation and subsequent evolution of a protostellar disk starting from a molecular cloud core, for times ranging up $0.5$~Myr. The disk,
fed by an infalling envelope, experiences gravitational instability. It develops large-scale spiral arms, portions of which fragment. The resulting density structures at six epochs
spanning 0.12--0.43 Myr after the formation of the protostar are transformed into NIR scattered light images and simulated ALMA dust continuum maps using a 3D Monte Carlo radiative transfer code. Our main conclusions are as follows.
\begin{enumerate}
\item As long as they are not obscured by an intervening
envelope, GI-induced spiral arms viewed at modest inclinations ($\lesssim45^\circ$) are visible at distances up to 1 kpc with
the current suite of NIR imaging instrumentation
(including Gemini/GPI, VLT/SPHERE, and Subaru/HiCIAO/SCExAO).
\item The spiral arms and streamers in four FU Ori objects recently revealed by Subaru \citep{liu16,liu16fuori} resemble GI-induced structures in our models, supporting the idea that FUors represent GI-unstable disks \citep[e.g.,][]{vorobyov05, vorobyov10burst, vorobyov15}.
\item Clumps formed by disk fragmentation have such small vertical scale heights that they are shadowed and appear as surface brightness depressions in NIR scattered light.
\item Both spiral arms and fragments in GI-unstable disks can be resolved and readily detected (by $\sim$10$\sigma$ for arms and by $\gtrsim 40 \sigma$ for fragments) in ALMA dust continuum observations of sources at 400 pc with an angular resolution of $0.1\arcsec$ and one-hour integration times. The minimum detectable fragment mass is $\sim$1 $\mj$ under these observing  conditions.
\end{enumerate}

Future work can focus on developing a 3D code that treats the 
hydrodynamics and radiative transfer of self-gravitating disks 
self-consistently, and on allowing dust and gas to slip 
past each other. Although our work is deficient in these
regards, it points robustly to the observability
of gravitational instability in protostellar
disks, given the powerful instrumentation available today.


\section*{Acknowledgments}

We thank the anonymous referee for constructive suggestions that improved the quality of the paper, and Jim Stone for insightful discussions. This project is partially supported by NASA through Hubble Fellowship grant HST-HF-51320.01-A awarded to R.D. by the Space Telescope Science Institute, which is operated by the Association of Universities for Research in Astronomy, Inc., for NASA, under contract NAS 5-26555. E.I.V. and Y.P. acknowledge partial support from the RFBR grant 14-02-00719. E.C. is grateful for support from NASA, the National Science Foundation, and Berkeley's Center for Integrative Planetary Science.


\section*{Appendix: Relation between the emergent radiative flux and
midplane temperature}

Let us consider a locally isothermal disk in vertical hydrostatic equilibrium. 
In the plane-parallel approximation, the thermal structure of the disk can be described
by the following system of radiative transfer moment equations:
\begin{eqnarray}
&&\frac{dF}{dz} = c\rho\kappa_P(B-E)  \label{m1}\\
&&\frac{c}{3} \frac{dE}{dz} = -\rho\kappa_R F \label{m2},
\end{eqnarray}
where $c$ is the speed of light, $\rho$ the gas volume density, 
$\kappa_P$ and $\kappa_R$ the Planck and Rosseland mean opacities,
$E$ the radiation energy density, $F$ the radiative
flux, $B = a T^4$ the  radiation energy density
in thermal equilibrium ($a=4\sigma/c$ the radiation constant,
$T$ the gas temperature). Equation~\eqref{m1} indicates that
the radiative flux $F$ depends on the difference between emission and
absorption in the vertical column of the disk. Equation~\eqref{m2} expresses the relation
between the radiative flux and radiation energy density in the Eddington 
approximation. 

This system of equations is closed with the following equation:
\begin{equation}
\frac{dF}{dz} = \rho S, \label{m3}
\end{equation}
which states that the radiative flux $F$ is actually produced by a non-radiative
heating source $\rho S$. Here, $S$ is defined as the heating rate per unit mass.
It is convenient to rewrite these equations using the integrated surface density from 
the midplane to a given vertical distance $z$, $\overline\Sigma(z) = \int\limits_{0}^{z}\rho(z^\prime)dz^\prime$.
We note that $\overline\Sigma(h)\equiv \Sigma_{1/2}\equiv \Sigma/2$ is the gas surface 
density from the midplane to the disk surface. 
The resulting equations take the following form:
\begin{eqnarray}
&&c\kappa_P(B-E) = S \label{n1} \\
&&\frac{c}{3} \frac{dE}{d\overline\Sigma} = -\kappa_R F \label{n2}\\
&&\frac{dF}{d\overline\Sigma} = S. \label{n3}
\end{eqnarray}
We now assume that $S$ is constant in the vertical direction, 
meaning that $\rho S$ is proportional to the mass in the 
vertical column of the disk. Given that the radiative flux is zero
at the midplane, the integration of equation~\eqref{n3} yields:
\begin{equation}
F=S\overline\Sigma. \label{nn3}
\end{equation}
After substituting equation~\eqref{nn3} into equation~\eqref{n2} and
integrating equation~\eqref{n2} from the midplane to the surface of the disk, we obtain:
\begin{equation}
E(\Sigma_{1/2}) = E(0) - \dfrac{3\kappa_R S}{2c} \Sigma_{1/2}^2,
\label{k1}
\end{equation}
where $E(0)$ and $E(\Sigma_{1/2})$ are the radiation energy densities
at the midplane and at the disk surface, correspondingly.
Now, let us adopt the following boundary condition at the disk
surface:
\begin{equation}
F(\Sigma_{1/2}) =\frac{1}{2} c E(\Sigma_{1/2}),
\label{bound}
\end{equation}
which assumes that radiation escapes from the disk surface isotropically.
Using equation~\eqref{nn3} we obtain:
\begin{equation}
S=\frac{cE(\Sigma_{1/2})}{2\Sigma_{1/2}}. \label{S}
\end{equation}
Finally, substituting $S$ in equation~\eqref{k1}, we obtain 
the relation between the radiation energy density at the disk surface surface and 
midplane:
\begin{equation}
E(\Sigma_{1/2}) = \frac{E(0)}{1+\dfrac{3}{4}\tau_{\rm R}},
\label{ES0}
\end{equation}
where $\tau_{\rm R} = \kappa_{\rm R} \Sigma_{1/2}$ is the the 
Rosseland optical depth from the midplane to the disk surface.
Using equations~\eqref{n1} and \eqref{S}, $E(0)$ can be rewritten in the following form:
\begin{equation}
E(0)= B(0) - \frac{E(\Sigma_{1/2})}{2\tau_{\rm P}}, \label{E0}
\end{equation}
where $\tau_P = \kappa_P \Sigma_{1/2}$ is the Planck optical
depth from the midplane to the disk surface and $B(0)$ the Planck radiation energy density 
at the midplane. 
Substituting equation~\eqref{E0} into equation~\eqref{ES0}, we obtain:
\begin{equation}
E(\Sigma_{1/2})= \frac{2\tau_{\rm P} }{1+2\tau_{\rm P} + \dfrac{3}{2}\tau_{\rm R} \tau_{\rm P}} B(0)
\end{equation}
Using the boundary condition~\eqref{bound} and noting that $\sigma = ca/4$
we finally obtain the relation between the midplane temperature and the radiative flux emerging
from the disk surface:
\begin{equation}
F(\Sigma_{1/2})= \frac{4\tau_{\rm P} \sigma \thydro^4 }{1+2\tau_{\rm P} + \dfrac{3}{2}\tau_{\rm R}
\tau_{\rm P}},
\end{equation}
where $\thydro$ is the midplane temperature in hydro simulations.


\clearpage


\begin{table}[]
\centering
\footnotesize
\caption{Models}
\begin{tabular}{cccccccccc}
\hline
Time & $\mdisk$ & $\mstar$ & $T_\star$ & $L_\star$ & $\dot{M}$          & $F_{\rm 0^\circ,FR}$ & $F_{\rm 0^\circ,ALMA}$  & $F_{\rm 45^\circ,FR}$ & $F_{\rm 45^\circ,ALMA}$  \\ \hline
Myr  & $\msun$  & $\msun$  & K         & $L_\odot$ & $\msun$ yr$^{-1}$  & mJy & mJy  & mJy & mJy      \\ \hline
0.12 & 0.135    & 0.366    & 4270      & 5.15      & $1.1\times10^{-6}$ & 152 & 112  & 136 & 98\\
0.19 & 0.190    & 0.463    & 4100      & 5.5       & $1.3\times10^{-6}$ & 154 & 122 & 131 & 104 \\
0.23 & 0.203    & 0.514    & 4050      & 5.2       & $2.0\times10^{-6}$ & 191 & 133 & 169 & 115 \\
0.28 & 0.209    & 0.571    & 3830      & 7.2       & $1.8\times10^{-6}$ & 176 & 130 & 153 & 110 \\
0.34 & 0.171    & 0.615    & 3800      & 4.2       & $8.0\times10^{-6}$ & 117 & 84 & 101 & 67 \\
0.43 & 0.179    & 0.655    & 3800      & 4         & $8.0\times10^{-6}$ & 107 & 71 & 99 & 65 \\ \hline
\label{tab:models}
\tablecomments{Properties of the models. $L_\star$ includes both the photospheric and accretion luminosities. $F_{\rm 0^\circ,FR}$ and $F_{\rm 45^\circ,FR}$ are the total spectral flux densities from full-resolution 1.3 mm (ALMA band 6) MCRT images at 0$^\circ$ and $45^\circ$ viewing inclinations, respectively. $F_{\rm 0^\circ,ALMA}$ and $F_{\rm 45^\circ,ALMA}$ are the corresponding total flux densities in simulated ALMA observations using array configuration \#19 (beam size $\sim0.1\arcsec$) and an integration time of one hour (see also Figure~\ref{fig:image_z1}). See Section~\ref{sec:results} for details.}
\end{tabular}
\end{table}

\end{document}